\documentclass[a4paper,11pt]{article}
\usepackage{graphicx,amssymb,amsmath,ascmac,amscd,graphicx,color,epic,array,subfigure}
\newcommand{\qed}{\hfill\hbox{\rule[-2pt]{6pt}{6pt}}}
\def\C{{\mathbb C}}
\def\Z{{\mathbb Z}}
\def\R{{\mathbb R}}

\def\P{{\mathbb P}}
\def\T{{\mathbb T}}
\def\H{{\mathbb H}}

\def\nn{\nonumber}

\def\DIS{\displaystyle}
\def\HLINE{\noalign{\hrule height1pt}}

\def\nn{{\nonumber}}

\newtheorem{theorem}{Theorem}
\newtheorem{corollary}{Corollary}

\newtheorem{proposition}{Proposition}

\newtheorem{definition}{Definition}

\begin{document}
\title{\textbf{On the addition formula for the tropical Hesse pencil}}
\author{Atsushi \textsc{Nobe}\\
Department of Mathematics, Faculty of Education, Chiba University,\\
1-33 Yayoi-cho Inage-ku, Chiba 263-8522, Japan}

\date{}

\maketitle

\begin{abstract}
We give the addition formula for the tropical Hesse pencil, which is the tropicalization of the Hesse pencil parametrized by the level-three theta functions, via those for the ultradiscrete theta functions.
The ultradiscrete theta functions are reduced from the level-three theta functions through the procedure of ultradiscretization by choosing their parameters appropriately.
The parametrization of the level-three theta functions firstly introduced in \cite{KKNT09} gives an explicit correspondence between the amoeba of the real part of the Hesse cubic curve and the tropical Hesse curve. 
Moreover, through the parametrization, we obtain the subtraction-free forms of the addition formulae for the level-three theta functions, which lead to the addition formula for the tropical Hesse pencil in terms of the ultradiscretization.
Using the parametrization, the tropical counterpart of the Hesse configuration is also given.
\end{abstract}

\section{Introduction}
In recent papers \cite{KNT08,KKNT09}, the author and his collaborators study several solvable chaotic dynamical systems given by piecewise linear maps.
The maps are arising from the duplication formulae for tropical elliptic pencils and are directly connected with those for elliptic pencils over $\C$ in terms of the procedure of ultradiscretization.
The general solutions to the dynamical systems are concretely constructed by using the ultradiscrete theta functions which parametrize the tropical elliptic pencils.
Each ultradiscrete theta function can be obtained as the ultradiscretization of the theta function which parametrizes the elliptic pencil over $\C$.
In particular, in \cite{KKNT09}, we introduce the level-three theta functions $\theta_0(z,\tau)$, $\theta_1(z,\tau)$, and $\theta_2(z,\tau)$ parametrizing the Hesse cubic curve and the series of their functional relations called the addition formulae.
A specialization of the variables in the addition formulae induces the duplication formula for the Hesse pencil, which gives the solvable chaotic dynamical system.
Applying the procedure of ultradiscretization to the level-three theta functions, we systematically obtain both the piecewise linear dynamical system possessing chaotic property and its general solution.
In this process, parametrization of the level-three theta functions with positive numbers $\varepsilon$ and $K$, one of which,  $\varepsilon$, vanishes in the limiting procedure, plays an important role.
The dynamical system thus obtained can naturally be regarded as the one arising from the duplication of points on the tropical Hesse pencil.
Thus, via the duplication formula for the level-three theta functions, we can connect the solvable dynamical system arising from the Hesse pencil with that from the tropical Hesse pencil.

In this paper, we give the addition formula for the points on the member of the tropical Hesse pencil.
The formula is obtained from that for the Hesse pencil over $\C$ upon application of the procedure of ultradiscretization to the level-three theta functions.
In contrast to the duplication formula, the addition formula is a combination of the ultradiscrete analogues of those for the Hesse pencil.
Since it is known that the addition of points on a tropical elliptic pencil gives the ultradiscrete QRT system \cite{N08}, we can construct both chaotic and integrable dynamical systems on the tropical Hesse pencil in analogy to those on the Hesse pencil.

\section{Tropical Hesse pencil}
\subsection{Hesse pencil}
\begin{figure}[htbp]
\centering
{
\includegraphics{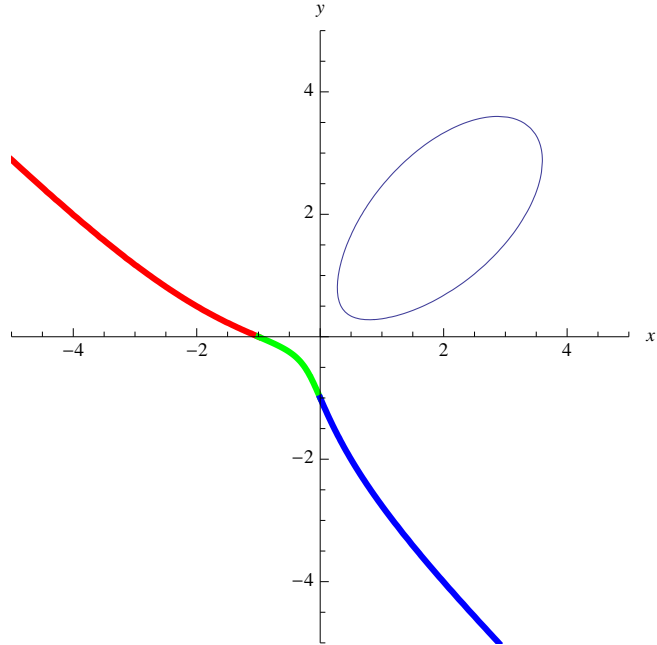}
}
\caption{The real part of the Hesse cubic curve. 
}
\label{fig:HesseCurve}
\end{figure}

The Hesse pencil is a one-dimensional linear system of plane cubic curves in $\P^2(\C)$ given by 
\begin{align*}
t_0\left(x_0^3+x_1^3+x_2^3\right)+t_1 x_0x_1x_2=0,
\end{align*}
where $(x_0,x_1,x_2)$ is the homogeneous coordinate of $\P^2(\C)$ and the parameter $(t_0,t_1)$ ranges over $\P^1(\C)$ \cite{AD06,Nakamura01}. 
The curve composing the pencil is called the Hesse cubic curve (see figure \ref{fig:HesseCurve}).

Each member of the pencil is denoted by $E_{t_0,t_1}$ and the pencil itself by $\left\{E_{t_0,t_1}\right\}_{(t_0,t_1)\in\P^1(\C)}$.
The nine  base points of the pencil are given as follows
\begin{align*}
&p_0=(0,1,-1)&
&p_1=(0,1,-\zeta_3)&
&p_2=(0,1,-\zeta_3^2)\\
&p_3=(1,0,-1)&
&p_4=(1,0,-\zeta_3^2)&
&p_5=(1,0,-\zeta_3)\\
&p_6=(1,-1,0)&
&p_7=(1,-\zeta_3,0)&
&p_8=(1,-\zeta_3^2,0),
\end{align*}
where $\zeta_3$ denotes the primitive third root of 1.

Any smooth curve in the pencil has the nine base points as its inflection points, and hence they are in the Hesse configuration \cite{AD06,SS31}.
The Hesse configuration is an arrangement of 9 points and 12 lines in the projective plane $\P^2(\C)$ which satisfies the following two conditions;
\begin{itemize}
\item each line passes through three of the 9 points and 
\item each point lies on four of the 12 lines.
\end{itemize}
Once an elliptic curve is given then its 9 inflection points and 12 inflection lines\footnote{A line passes through three inflection points is called an inflection line.} realize the Hesse configuration.
In particular, all non-singular members in the Hess pencil have the 9 inflection points $p_0,p_1,\cdots,p_8$ and the 12 inflection lines in common, hence each of them has the unique realization of the Hesse configuration.
Note that the 12 inflection lines are the irreducible components of the singular members $E_{0,1}$, $E_{1,-3}$, $E_{1,-3\zeta_3}$, and $E_{1,-\zeta_3^2}$ of the pencil given below (see (\ref{eq:sinchp1} -- \ref{eq:sinchp4})).


\begin{table}[htbp]
\renewcommand{\arraystretch}{1.2}
\begin{center}
\caption{The inflection lines and the inflection points in the Hesse configuration.}
\begin{tabular}{clc}\label{tab:Hesseconfig}
\\\HLINE
Singular curves&Inflection lines&Inflection points\\\hline
$E_{0,1}$&
$\left\{\begin{array}{l}x_0=0\\x_1=0\\x_2=0\end{array}\right.$&
$\left.\begin{array}{l}p_0,p_1,p_2\\ p_3,p_4,p_5\\ p_6,p_7,p_8\end{array}\right.$\\
$E_{1,-3}$&
$\left\{\begin{array}{l}x_0+x_1+x_2=0\\x_0+\zeta_3x_1+\zeta_3^2x_2=0\\x_0+\zeta_3^2x_1+\zeta_3x_2=0\end{array}\right.$&
$\left.\begin{array}{l}p_0,p_3,p_6\\ p_2,p_5,p_8\\ p_1,p_4,p_7\end{array}\right.$\\
$E_{1,-3\zeta_3}$&
$\left\{\begin{array}{l}x_0+\zeta_3x_1+x_2=0\\x_0+\zeta_3^2x_1+\zeta_3^2x_2=0\\x_0+x_1+\zeta_3x_2=0\end{array}\right.$&
$\left.\begin{array}{l}p_1,p_3,p_8\\ p_0,p_5,p_7\\ p_2,p_4,p_6\end{array}\right.$\\
$E_{1,-3\zeta_3^2}$&
$\left\{\begin{array}{l}x_0+\zeta_3^2x_1+x_2=0\\x_0+\zeta_3x_1+\zeta_3x_2=0\\x_0+x_1+\zeta_3^2x_2=0\end{array}\right.$&
$\left.\begin{array}{l}p_2,p_3,p_7\\ p_0,p_4,p_8\\ p_1,p_5,p_6\end{array}\right.$\\
\HLINE
\end{tabular}
\end{center}
\end{table}

The Weierstra$\ss$ form of the Hesse cubic curve is given as follows
\begin{align*}
x_1^{\prime 2}x_2^\prime
=
x_0^{\prime 3}+A(t_0,t_1)x_0^\prime x_2^{\prime 2}+B(t_0,t_1)x_2^{\prime 3},
\end{align*}
where $(x_0^\prime,x_1^\prime,x_2^\prime)\in\P^2(\C)$ is the homogeneous coordinate of $\P^2(\C)$, $(t_0,t_1)=(u_0,6u_1)$, and
\begin{align*}
&
A(t_0,t_1)
=
12u_1\left(u_0^3-u_1^3\right)\\
&
B(t_0,t_1)
=
2\left(u_0^6-20u_0^3u_1^3-8u_1^6\right).
\end{align*}
The transformation ${}^t(x_0,x_1,x_2)\mapsto{}^t(x_0^\prime,x_1^\prime,x_2^\prime)$, form the Hesse form to the Weierstra$\ss$ form, is given by the linear map
\begin{align*}
\left(\begin{matrix}
9t_0t_1^2&9t_0t_1^2&108t_0^3+t_1^3\\
-2\sqrt{6}it_0(27t_0^3+t_1^3)&2\sqrt{6}it_0(27t_0^3+t_1^3)&0\\
54t_0&54t_0&-18t_1\\
\end{matrix}\right).
\end{align*} 
The discriminant of the Weierstra$\ss$ cubic curve is 
\begin{align*}
\Delta
=
4A^3+27B^2
=
2^23^3u_0^3\left(u_0^3+2^3u_1^3\right).
\end{align*}
Thus we see that the singular member of the Hesse pencil is described by
\begin{align*}
(t_0,t_1)
=
(0,1),\
(1,-3),\
(1,-3\zeta_3),\
(1,-3\zeta_3^2),
\end{align*}
or explicitly given by the unions of three lines:
\begin{align}
&E_{0,1}:&
x_0x_1x_2=0&
\label{eq:sinchp1}\\
&E_{1,-3}:&
\left(x_0+x_1+x_2\right)\left(x_0+\zeta_3x_1+\zeta_3^2x_2\right)\left(x_0+\zeta_3^2x_1+\zeta_3x_2\right)=0&
\label{eq:sinchp2}\\
&E_{1,-3\zeta_3}:&
\left(x_0+\zeta_3x_1+x_2\right)\left(x_0+\zeta_3^2x_1+\zeta_3^2x_2\right)\left(x_0+x_1+\zeta_3x_2\right)=0&
\label{eq:sinchp3}\\
&E_{1,-3\zeta_3^2}:&
\left(x_0+\zeta_3^2x_1+x_2\right)\left(x_0+\zeta_3x_1+\zeta_3x_2\right)\left(x_0+x_1+\zeta_3^2x_2\right)=0&.
\label{eq:sinchp4}
\end{align}
Note that each singular member has multiplicity three.
Table \ref{tab:Hesseconfig} shows the inflection lines and the inflection points in the Hesse configuration.


\subsection{Tropicalization}
Let us consider tropicalization of the Hesse pencil.
For the defining polynomial of the Hesse cubic curve
\begin{align*}
f(x_0,x_1,x_2; t_0,t_1)
:=
t_0\left(x_0^3+x_1^3+x_2^3\right)+t_1 x_0x_1x_2,
\end{align*}
we apply the procedure of tropicalization.
At first, replace the addition $+$ and the multiplication $\times$ with the tropical addition $\oplus$ and the tropical multiplication $\otimes$ respectively; then we obtain the {tropical} polynomial
\begin{align*}
\tilde f(\tilde x_0,\tilde x_1,\tilde x_2; \tilde t_0,\tilde t_1)
:=
\tilde t_0\otimes \left(\tilde x_0^{\otimes3}\oplus \tilde x_1^{\otimes3}\oplus \tilde x_2^{\otimes3}\right)
\oplus 
\tilde t_1\otimes \tilde x_0\otimes \tilde x_1\otimes \tilde x_2.
\end{align*}
In order to distinguish tropical variables form original ones, we ornament them with $\tilde{}$.
The tropical operations $\oplus$ and $\otimes$ are defined as follows
\begin{align*}
a\oplus b
=
\max(a,b)
\qquad
a\otimes b
=
a+b
\qquad
\mbox{for $a,b\in \T:=\R\cup\{-\infty\}$},
\end{align*}
where $\T$ is the tropical semi-field.
Thus the tropical polynomial $\tilde f(\tilde x_0,\tilde x_1,\tilde x_2; \tilde t_0,\tilde t_1)$ reduces to
\begin{align*}
\tilde f(\tilde x_0,\tilde x_1,\tilde x_2; \tilde t_0,\tilde t_1)
=
\max\left(
\tilde t_0+3\tilde x_0,\tilde t_0+3\tilde x_1,\tilde t_0+3\tilde x_2,\tilde t_1+\tilde x_0+\tilde x_1+\tilde x_2
\right).
\end{align*}

Noting
\begin{align*}
a\oplus (-\infty)
=
a
\qquad
a\otimes0
=
a
\qquad
a\otimes(-a)
=
0,
\end{align*}
we see that $-\infty$ and $0$ are the units of addition and multiplication in $\T$ respectively.
We find the element $-a$ as the inverse of $a$ with respect to the multiplication, however, there is no inverse of $a$ with respect to the addition. 

\begin{definition}\normalfont(See definition 1.1 in \cite{MZ06})
The tropical projective space $\P^{n,trop}$ consists of the classes of $(n+1)$-tuples $(x_0,\cdots,x_n)\in\T^{n+1}$ such that not all of them are equal to $-\infty$ with respect to the following equivalence relation $\sim$;
\begin{align*}
(x_0,\cdots,x_n)
\sim
(y_0,\cdots,y_n)
\quad
&\Longleftrightarrow
\quad
x_i=y_i=\lambda
\quad
\mbox{for $i=0,\ldots,n$ and $\lambda\in\R$},
\end{align*}
where we assume $x_0\otimes\cdots\otimes x_n\neq-\infty$ and $y_0\otimes\cdots\otimes y_n\neq-\infty$.
\end{definition}

Under the identification $(x_1, . . . , x_n)\in\R^n$ with $(0,x_1, . . . , x_n)\in\P^{n,trop}$ the real space $\R^n$ is contained in $\P^{n,trop}$. 
Thus we have an embedding
$\iota_n:
\R^n\subset \P^{n,trop}$.
Also we have $n + 1$ affine charts $\T^n\to\P^{n,trop}$, given by the (tropical) ratio with the $j$-th coordinate.

Let $(t_0,t_1)$ be a point in $\P^{1,trop}$.
Then $\tilde f$ can be regarded as a function $\tilde f:\P^{2,trop}\to\T$.
The tropical Hesse curve is the set of points such that the function $\tilde f$ is not differentiable. 
We denote the tropical Hesse curve by $C_{\tilde{t}_0,\tilde{t}_1}$.
Upon introduction of the inhomogeneous coordinate $(X:=\tilde{x_1}-\tilde{x_0},Y:=\tilde{x_2}-\tilde{x_0})\in\P^{2,trop}$ and $K:=\tilde{t}_1-\tilde{t}_0\in\P^{1,trop}$ the tropical Hesse curve is denoted by $C_K$ and is given by the tropical polynomial
\begin{align*}
F(X,Y; K)
:=
\max\left(
3X,3Y,0,K+X+Y
\right).
\end{align*}

\begin{figure}[htbp]
\centering
{\unitlength=.05in{\def\arraystretch{1.0}
\begin{picture}(50,65)(-25,-30)
\thicklines
\put(0,28){\vector(0,1){2}}
\dottedline(0,-30)(0,28)
\put(0,32){\makebox(0,0){$Y$}}
\put(28,0){\vector(1,0){2}}
\dottedline(-30,0)(28,0)
\put(32,0){\makebox(0,0){$X$}}
\linethickness{1.6pt}
\dashline[10]{5}(0,0.1)(30,30.1)
\dashline[10]{5}(0,-0.1)(30,29.9)
\dashline[10]{5}(0,0)(0,-30)
\dashline[10]{5}(0,0)(-30,0)
\thicklines
\put(-25,0){\line(1,-1){25}}
\put(0,-25){\line(1,2){25}}
\put(-25,0){\line(2,1){50}}
\put(-20,0){\line(1,-1){20}}
\put(0,-20){\line(1,2){20}}
\put(-20,0){\line(2,1){40}}
\put(-10,0){\line(1,-1){10}}
\put(0,-10){\line(1,2){10}}
\put(-10,0){\line(2,1){20}}
\put(-10,0){\line(-1,0){20}}
\put(0,-10){\line(0,-1){20}}
\put(10,10){\line(1,1){20}}
\put(-3,-3){\makebox(0,0){$O$}}
\end{picture}
}}
\caption{Several members of the tropical Hesse pencil. The ones drawn with solid lines are the regular members and the one with broken line is the singular member for $K=0$.}
\label{fig:tropHesse}
\end{figure}
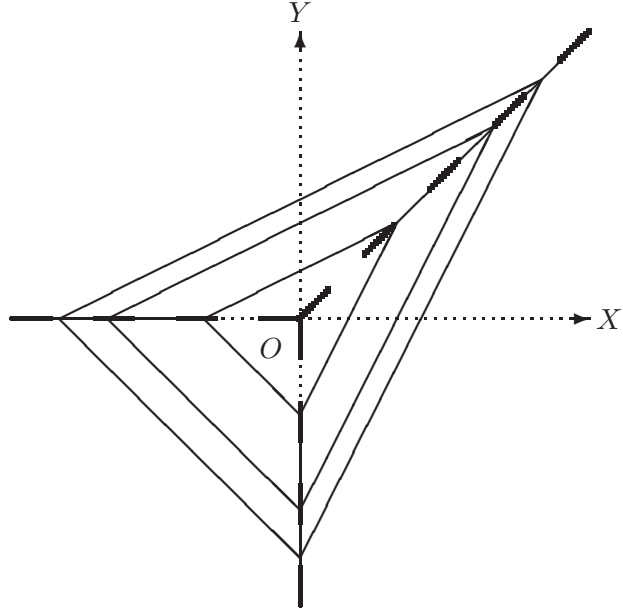

Figure \ref{fig:tropHesse} shows the tropical Hesse curves.
The one-dimensional linear system $\{C_K\}_{K\in\P^{1,trop}}$ consisting of the tropical Hesse curves is called the tropical Hesse pencil.
The complement of the tentacles, i.e., the finite part, of $C_K$ is denoted by $\bar C_K$. 
We denote the vertices whose coordinates are $(K,K)$, $(-K,0)$, and $(0,-K)$ by $V_1$, $V_2$, and $V_3$, respectively.
Also denote the edges $\overline{V_1V_2}$, $\overline{V_1V_2}$, and $\overline{V_1V_2}$ by $E_1$, $E_2$, and $E_3$, respectively.

Each member $C_K$ of the tropical Hesse pencil for generic choice of $K\in\P^{1,trop}$ has genus one, therefore it can be regarded as a tropical elliptic pencil.
Singular curves appear only for the choice of the parameter as $K=\infty$ or $K\leq0$.
For $K=\infty$, or equivalently for $(\tilde{t}_0,\tilde{t}_1)=(-\infty,0)$, the tropical polynomial $\tilde f(\tilde x_0,\tilde x_1,\tilde x_2; \tilde t_0,\tilde t_1)$ reduces to
\begin{align}
\tilde f(\tilde x_0,\tilde x_1,\tilde x_2; -\infty,0)
=
\tilde x_0\otimes\tilde x_1\otimes\tilde x_2
=
\tilde x_0+\tilde x_1+\tilde x_2.
\label{eq:defpolysing1}
\end{align}
This can not be differentiated\footnote{For example, for fixed $\tilde{x}_1,\tilde{x}_2\in\T$ and $h>0$, the difference $$\left\{\tilde f(-\infty+h,\tilde x_1,\tilde x_2; -\infty,0)-\tilde f(-\infty,\tilde x_1,\tilde x_2; -\infty,0)\right\}/h$$ along with $\tilde{x}_0$ can not be defined.} at the boundary of  $\P^{2,trop}$, therefore the curve $C_K$ reduces to the union of three tropical lines which compose the boundary of $\P^{2,trop}$:
\begin{align}
C_\infty:\quad
\{\tilde x_0=-\infty\}\cup
\{\tilde x_1=-\infty\}\cup
\{\tilde x_2=-\infty\}.
\label{eq:cinf}
\end{align}
Since the defining polynomial of the singular curve $E_{0,1}$ is
\begin{align}
f(x_0,x_1,x_2; 0,1)
=
x_0x_1x_2,
\label{eq:singpoly1}
\end{align}
the tropical polynomial \eqref{eq:defpolysing1} can be regarded as the tropicalization of \eqref{eq:singpoly1}.
Therefore, the singular curve $C_\infty$ is the tropical counterpart of $E_{0,1}$. 

On the other hand, for $K\leq0$, or equivalently for $\tilde{t}_0\geq \tilde{t}_1$, $F(X,Y;K)$ reduces to
\begin{align}
F(X,Y; K)
=
3\max\left(
X,Y,0
\right),
\label{eq:singtroppoly2}
\end{align}
or equivalently $\tilde f(\tilde x_0,\tilde x_1,\tilde x_2; \tilde t_0,\tilde t_1)$ to
\begin{align}
\tilde f(\tilde x_0,\tilde x_1,\tilde x_2; \tilde t_0,\tilde t_1)
=
\left(\tilde x_0\oplus \tilde x_1\oplus \tilde x_2\right)^{\otimes3}
=
3\max\left(\tilde x_0,\tilde x_1,\tilde x_2\right).
\label{eq:defpolysing2}
\end{align}
The tropical curve given by \eqref{eq:singtroppoly2} is clearly independent of $K$.
Hence we take $C_0$ as the representative of the singular curves $C_K$ for $K\leq0$. 
The curve $C_0$ is a triple tropical line whose only vertex is on the origin (see figure \ref{fig:tropHesse}).
The tropical polynomial \eqref{eq:defpolysing2} can be regarded as the tropicalization of the polynomials in \eqref{eq:sinchp2}, \eqref{eq:sinchp3}, and \eqref{eq:sinchp4}, which give the singular curves $E_{1,-3}$, $E_{1,-3\zeta_3}$, and $E_{1,-3\zeta_3^2}$, respectively.
Thus the curve $C_0$ is regarded as the tropical counterpart of $E_{1,-3}$, $E_{1,-3\zeta_3}$, and $E_{1,-3\zeta_3^2}$.
Table \ref{tab:udsingcurve} shows the correspondence between the defining polynomials of the singular members of the Hesse pencil and their tropical counterparts.
\begin{table}[htbp]
\renewcommand{\arraystretch}{1.2}
\begin{center}
\caption{The correspondence between the defining polynomials of the singular members of the Hesse pencil and their tropical counterparts.}
\begin{tabular}{clrc}\label{tab:udsingcurve}
\\\HLINE
\multicolumn{2}{c}{Hesse pencil}&
\multicolumn{2}{c}{Tropical Hesse pencil}\\
Singular curves&Inflection lines&Inflection lines&Singular curves\\\hline
$E_{0,1}$&$\left\{\begin{array}{l}x_0\\x_1\\x_2\end{array}\right.$&$\left.\begin{array}{l}\tilde x_0\\\tilde x_1\\\tilde x_2\end{array}\right\}$&$C_\infty$\\\hline
$E_{1,-3}$&$\left\{\begin{array}{l}x_0+x_1+x_2\\x_0+\zeta_3x_1+\zeta_3^2x_2\\x_0+\zeta_3^2x_1+\zeta_3x_2\end{array}\right.$&&\\
$E_{1,-3\zeta_3}$&$\left\{\begin{array}{l}x_0+\zeta_3x_1+x_2\\x_0+\zeta_3^2x_1+\zeta_3^2x_2\\x_0+x_1+\zeta_3x_2\end{array}\right.$&
$\left.\begin{array}{l}\max\left(\tilde x_0, \tilde x_1, \tilde x_2\right)\\\max\left(\tilde x_0, \tilde x_1, \tilde x_2\right)\\\max\left(\tilde x_0, \tilde x_1, \tilde x_2\right)\end{array}\right\}$&$C_0$\\
$E_{1,-3\zeta_3^2}$&$\left\{\begin{array}{l}x_0+\zeta_3^2x_1+x_2\\x_0+\zeta_3x_1+\zeta_3x_2\\x_0+x_1+\zeta_3^2x_2\end{array}\right.$&&\\
\HLINE
\end{tabular}
\end{center}
\end{table}

Vigeland showed that a tropical elliptic curve has an additive group structure in analogy to an elliptic curve \cite{Vigeland04}.
The group structure is induced from that of the Jacobian of the tropical elliptic curve, which is isomorphic to $S^1$, to its complement of the tentacles via the Abel-Jacobi map.
Therefore we have the group isomorphism
\begin{align*}
\bar C_K\simeq C_K/\sim\ {\longrightarrow}\ J(C_K)\simeq S^1,
\end{align*}
where $\sim$ is an equivalence relation called the linear equivalence \cite{Vigeland04}.
In the following, we give an explicit formula for the addition of points on the tropical Hesse curve via the ultradiscretization of those for the level-three theta functions. 

\section{Level-three theta functions}
\subsection{Definition}
The level-three theta functions $\theta_0(z,\tau)$, $\theta_1(z,\tau)$, and $\theta_2(z,\tau)$ are defined by using the theta function $\vartheta_{(a,b)}(z,\tau)$ with characteristics:
\begin{align*}
\theta_{k}(z,\tau)
:=
\vartheta_{\left(\frac{k}{3}-\frac{1}{6},\frac{3}{2}\right)}(3z,3\tau)
=
\sum_{n\in\Z}e^{3\pi i\left(n+\frac{k}{3}-\frac{1}{6}\right)^{2}\tau}
e^{6\pi i\left(n+\frac{k}{3}-\frac{1}{6}\right)\left(z+\frac{1}{2}\right)}
\qquad(k=0,1,2),
\end{align*}
where $z\in\C$ and $\tau\in\H:=\{\tau\in\C\ |\ {\rm Im}\mkern2mu \tau>0\}$.

Fix $\tau\in\H$. 
For simplicity, we abbreviate $\theta_k(z,\tau)$ and $\theta_k(0,\tau)$ as $\theta_k(z)$ and $\theta_k$ for $k=0,1,2$, respectively.
The level-three theta functions have the quasi periodicity \cite{KKNT09}
\begin{align}
&\theta_k(z+1)
=
-\theta_k(z)
\label{eq:thetaprop1}\\
&\theta_k(z+\tau)
=
-e^{3\pi i \tau}e^{-6\pi iz}\theta_k(z)
\label{eq:thetaprop2}
\end{align}
for $k=0,1,2$.
Let $L_\tau:=(-\tau)\Z+(3\tau+1)\Z$ be a lattice in $\C$.
Noting
\begin{align*}
\left(
\begin{matrix}
-1&3\\
0&1\\
\end{matrix}
\right)^{-1}
=
\left(
\begin{matrix}
-1&3\\
0&1\\
\end{matrix}
\right),
\end{align*}
we have an isomorphism $L_\tau\simeq\C/\Z+\tau\Z$.
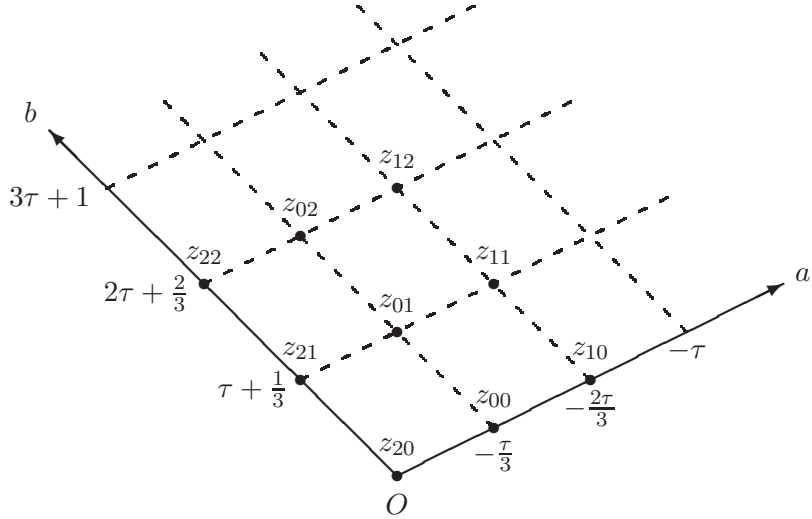
\begin{figure}[htbp]
\centering
{\unitlength=.05in{\def\arraystretch{1.0}
\begin{picture}(50,55)(-25,-3)
\thicklines

\put(0,0){\vector(-1,1){36}}
\put(-38,38){\makebox(0,0){$b$}}
\put(0,0){\vector(2,1){40}}
\put(42,21){\makebox(0,0){$a$}}

\dashline[10]{1}(-10,10)(28,29)
\dashline[10]{1}(-20,20)(18,39)
\dashline[10]{1}(-30,30)(8,49)
\dashline[10]{1}(10,5)(-24,39)
\dashline[10]{1}(20,10)(-14,44)
\dashline[10]{1}(30,15)(-4,49)

\put(0,-3){\makebox(0,0){$O$}}
\put(0,3){\makebox(0,0){$z_{20}$}}
\put(10,2){\makebox(0,0){$-\frac{\tau}{3}$}}
\put(10,8){\makebox(0,0){$z_{00}$}}
\put(20,7){\makebox(0,0){$-\frac{2\tau}{3}$}}
\put(20,13){\makebox(0,0){$z_{10}$}}
\put(30,13){\makebox(0,0){$-\tau$}}
\put(-15,9){\makebox(0,0){$\tau+\frac{1}{3}$}}
\put(-10,13){\makebox(0,0){$z_{21}$}}
\put(-26,19){\makebox(0,0){$2\tau+\frac{2}{3}$}}
\put(-20,23){\makebox(0,0){$z_{22}$}}
\put(-36,29){\makebox(0,0){$3\tau+1$}}

\put(0,18){\makebox(0,0){$z_{01}$}}
\put(-10,28){\makebox(0,0){$z_{02}$}}

\put(10,23){\makebox(0,0){$z_{11}$}}
\put(0,33){\makebox(0,0){$z_{12}$}}

\put(0,0){\circle*{1}}
\put(-10,10){\circle*{1}}
\put(-20,20){\circle*{1}}
\put(10,5){\circle*{1}}
\put(0,15){\circle*{1}}
\put(-10,25){\circle*{1}}
\put(20,10){\circle*{1}}
\put(10,20){\circle*{1}}
\put(0,30){\circle*{1}}
\end{picture}
}}
\caption{The zeros of $\theta_k(z,\tau)$ for $k=0,1,2$ in the fundamental domain.}
\label{fig:lattice}
\end{figure}
Let us denote the axes in the directions $-\tau$ and $3\tau+1$ by $a$ and $b$ respectively (see figure \ref{fig:lattice}).

\subsection{Addition formulae}
\begin{theorem}\normalfont
\label{thm:addfltt}
For a fixed $\tau\in\H$, the level-three theta functions $\theta_0(z,\tau)$, $\theta_1(z,\tau)$, and $\theta_2(z,\tau)$ satisfy the following 9 functional relations called the addition formulae \cite{KKNT09}
\begin{subequations}
\begin{align}
\theta_{0}^{2}\theta_{0}(z+w)\theta_{0}(z-w)
&=
\theta_{1}(z)\theta_{2}(z)\theta_{2}(w)^{2}
-
\theta_{0}(z)^{2}\theta_{0}(w)\theta_{1}(w)
\label{eq:addtheta1a}\\
\theta_{0}^{2}\theta_{1}(z+w)\theta_{0}(z-w)
&=
\theta_{0}(z)\theta_{1}(z)\theta_{1}(w)^{2}
-
\theta_{2}(z)^{2}\theta_{0}(w)\theta_{2}(w)
\label{eq:addtheta1b}\\
\theta_{0}^{2}\theta_{2}(z+w)\theta_{0}(z-w)
&=
\theta_{0}(z)\theta_{2}(z)\theta_{0}(w)^{2}
-
\theta_{1}(z)^{2}\theta_{1}(w)\theta_{2}(w)
\label{eq:addtheta1c}
\end{align}
\end{subequations}
\begin{subequations}
\begin{align}
\theta_{0}^{2}\theta_{0}(z+w)\theta_{1}(z-w)
&=
\theta_{0}(z)\theta_{1}(z)\theta_{0}(w)^{2}
-
\theta_{2}(z)^{2}\theta_{1}(w)\theta_{2}(w)
\label{eq:addtheta2a}\\
\theta_{0}^{2}\theta_{1}(z+w)\theta_{1}(z-w)
&=
\theta_{0}(z)\theta_{2}(z)\theta_{2}(w)^{2}
-
\theta_{1}(z)^{2}\theta_{0}(w)\theta_{1}(w)
\label{eq:addtheta2b}\\
\theta_{0}^{2}\theta_{2}(z+w)\theta_{1}(z-w)
&=
\theta_{1}(z)\theta_{2}(z)\theta_{1}(w)^{2}
-
\theta_{0}(z)^{2}\theta_{0}(w)\theta_{2}(w)
\label{eq:addtheta2c}
\end{align}
\end{subequations}
\begin{subequations}
\begin{align}
\theta_{0}^{2}\theta_{0}(z+w)\theta_{2}(z-w)
&=
\theta_{0}(z)\theta_{2}(z)\theta_{1}(w)^{2}
-
\theta_{1}(z)^{2}\theta_{0}(w)\theta_{2}(w)
\label{eq:addtheta3a}\\
\theta_{0}^{2}\theta_{1}(z+w)\theta_{2}(z-w)
&=
\theta_{1}(z)\theta_{2}(z)\theta_{0}(w)^{2}
-
\theta_{0}(z)^{2}\theta_{1}(w)\theta_{2}(w)
\label{eq:addtheta3b}\\
\theta_{0}^{2}\theta_{2}(z+w)\theta_{2}(z-w)
&=
\theta_{0}(z)\theta_{1}(z)\theta_{2}(w)^{2}
-
\theta_{2}(z)^{2}\theta_{0}(w)\theta_{1}(w),
\label{eq:addtheta3c}
\end{align}
\end{subequations}
where $z,w\in\C$. 
\end{theorem}

It follows from theorem \ref{thm:addfltt} that we have \cite{KKNT09}
\begin{align}
&{\theta_2^\prime}\left(\theta_0(z)^3+\theta_1(z)^3+\theta_2(z)^3\right)+6{\theta_0^\prime}\theta_0(z)\theta_1(z)\theta_2(z)=0.
\label{eq:thetaprop3}
\end{align}
Consider a map $\varphi:\C\to\P^2(\C)$,
\begin{align*}
\varphi:\ z\longmapsto (\theta_2(z),\theta_0(z),\theta_1(z)).
\end{align*}
This induces a map from the complex torus $\C/L_\tau$ to the Hesse cubic curve $E_{{\theta_2^\prime},6{\theta_0^\prime}}$ due to \eqref{eq:thetaprop1}, \eqref{eq:thetaprop2}, and \eqref{eq:thetaprop3}.
This map is known to give an isomorphism $\C/L_\tau\simeq E_{{\theta_2^\prime},6{\theta_0^\prime}}$.
Thus the level-three theta functions parametrize the Hesse cubic curve.

Considering (\ref{eq:addtheta1a} -- \ref{eq:addtheta1c}), the point $(\theta_2(z+w),\theta_0(z+w),\theta_1(z+w))$ is computed as follows
\begin{align}
(\theta_2(z+w),\theta_0(z+w),\theta_1(z+w))
=
(
&
\theta_{0}(z)\theta_{2}(z)\theta_{0}(w)^{2}
-
\theta_{1}(z)^{2}\theta_{1}(w)\theta_{2}(w),\nn\\
&\qquad
\theta_{1}(z)\theta_{2}(z)\theta_{2}(w)^{2}
-
\theta_{0}(z)^{2}\theta_{0}(w)\theta_{1}(w),\nn\\
&\qquad\qquad
\theta_{0}(z)\theta_{1}(z)\theta_{1}(w)^{2}
-
\theta_{2}(z)^{2}\theta_{0}(w)\theta_{2}(w)
)
\label{eq:addtheta1}
\end{align}
except for $z,w\in\C/L_\tau$ satisfying $\theta_0(z-w)=0$. 
Similarly, considering (\ref{eq:addtheta2a} -- \ref{eq:addtheta2c}) and (\ref{eq:addtheta3a} -- \ref{eq:addtheta3c}), we obtain the following
\begin{align}
(\theta_2(z+w),\theta_0(z+w),\theta_1(z+w))
=
(
&
\theta_{1}(z)\theta_{2}(z)\theta_{1}(w)^{2}
-
\theta_{0}(z)^{2}\theta_{0}(w)\theta_{2}(w),\nn\\
&\qquad
\theta_{0}(z)\theta_{1}(z)\theta_{0}(w)^{2}
-
\theta_{2}(z)^{2}\theta_{1}(w)\theta_{2}(w),\nn\\
&\qquad\qquad
\theta_{0}(z)\theta_{2}(z)\theta_{2}(w)^{2}
-
\theta_{1}(z)^{2}\theta_{0}(w)\theta_{1}(w)
)
\label{eq:addtheta2}\\
(\theta_2(z+w),\theta_0(z+w),\theta_1(z+w))
=
(
&
\theta_{0}(z)\theta_{1}(z)\theta_{2}(w)^{2}
-
\theta_{2}(z)^{2}\theta_{0}(w)\theta_{1}(w),\nn\\
&\qquad
\theta_{0}(z)\theta_{2}(z)\theta_{1}(w)^{2}
-
\theta_{1}(z)^{2}\theta_{0}(w)\theta_{2}(w),\nn\\
&\qquad\qquad
\theta_{1}(z)\theta_{2}(z)\theta_{0}(w)^{2}
-
\theta_{0}(z)^{2}\theta_{1}(w)\theta_{2}(w)
)
\label{eq:addtheta3}
\end{align}
except for $z,w\in\C/L_\tau$ satisfying $\theta_1(z-w)=0$ and $\theta_2(z-w)=0$, respectively. 
Since the zeros of $\theta_0(z)$, $\theta_1(z)$, and $\theta_2(z)$ never coincide with each other, at least two of the addition formulae (\ref{eq:addtheta1} -- \ref{eq:addtheta3}) can be defined for any $z,w\in\C/L_\tau$.
Moreover, by using the relation \eqref{eq:thetaprop3}, we can prove that the three formulae (\ref{eq:addtheta1} -- \ref{eq:addtheta3}) are essentially the same where they are defined simultaneously.
Thus the addition formula for the Hesse cubic curve is uniquely defined on $\C/L_\tau$.

The isomorphism $\varphi: \C/L_\tau\to E_{{\theta_2^\prime},6{\theta_0^\prime}}$ induces the additive group structure on $E_{{\theta_2^\prime},6{\theta_0^\prime}}$ from $\C/L_\tau$ through the addition formulae for the level-three theta functions.
The relation \eqref{eq:thetaprop4} (see below) implies
\begin{align*}
\varphi:\ 0\longmapsto (\theta_2,\theta_0,\theta_1)=(0,1,-1)=p_0.
\end{align*}
Thus we obtain the addition formulae for the Hesse cubic curve $(E_{{\theta_2^\prime},6{\theta_0^\prime}},p_0)$ equipped with the unit of addition $p_0$.
\begin{theorem}\normalfont
Let the unit of addition on the Hesse cubic curve $E_{{\theta_2^\prime},6{\theta_0^\prime}}$ be $p_0=(0,1,-1)$.
Let $(x_0,x_1,x_2)$ and $(x_0^\prime,x_1^\prime,x_2^\prime)$ be points on $E_{{\theta_2^\prime},6{\theta_0^\prime}}$.
Then the addition $(x_0,x_1,x_2)+(x_0^\prime,x_1^\prime,x_2^\prime)$ of the points is given as follows
\begin{align*}
(x_0,x_1,x_2)+(x_0^\prime,x_1^\prime,x_2^\prime)
&=
(
x_1x_2{x_2^\prime}^2-x_0^2x_0^\prime x_1^\prime,
x_0x_1{x_1^\prime}^2-x_2^2x_0^\prime x_2^\prime,
x_0x_2{x_0^\prime}^2-x_1^2x_1^\prime x_2^\prime
)\\
&=
(
x_0x_1{x_0^\prime}^2-x_2^2x_1^\prime x_2^\prime,
x_0x_2{x_2^\prime}^2-x_1^2x_0^\prime x_1^\prime,
x_1x_2{x_1^\prime}^2-x_0^2x_0^\prime x_2^\prime
)\\
&=
(
x_0x_2{x_1^\prime}^2-x_1^2x_0^\prime x_2^\prime,
x_1x_2{x_0^\prime}^2-x_0^2x_1^\prime x_2^\prime,
x_0x_1{x_2^\prime}^2-x_2^2x_0^\prime x_1^\prime
).
\end{align*}
\end{theorem}

We can easily see that the following property holds
\begin{align}
&\theta_0=-\theta_1
\qquad\theta_2=0
\label{eq:thetaprop4}\\
&\theta_k\left(z+\frac{\tau}{3}\right)
=
-e^{\frac{1}{3}\pi i \tau}e^{-2\pi iz}\theta_{k+1}(z)
\label{eq:thetaprop5}\\
&\theta_k\left(z+\frac{1}{3}\right)
=
e^{2\pi i \left(\frac{k}{3}-\frac{1}{6}\right)}\theta_k(z).
\label{eq:thetaprop6}
\end{align}
The relations (\ref{eq:thetaprop4} -- \ref{eq:thetaprop6}) imply that we can take the following representatives $z_{0k},z_{k1},z_{k2}$ of the zeros of $\theta_k(z)$ in $\C/L_\tau$ for $k=0,1,2$ (see figure \ref{fig:lattice})
\begin{align*}
\left(
\begin{matrix}
z_{20}&z_{21}&z_{22}\\[5pt]
z_{00}&z_{01}&z_{02}\\[5pt]
z_{10}&z_{11}&z_{12}\\[5pt]
\end{matrix}
\right)
=
\left(
\begin{matrix}
0&\tau+\frac{1}{3}&2\tau+\frac{2}{3}\\[5pt]
-\frac{\tau}{3}&\frac{2\tau}{3}+\frac{1}{3}&\frac{5\tau}{3}+\frac{2}{3}\\[5pt]
-\frac{2\tau}{3}&\frac{\tau}{3}+\frac{1}{3}&\frac{4\tau}{3}+\frac{2}{3}\\[5pt]
\end{matrix}
\right).
\end{align*}
These nine zeros are mapped into the nine inflection points on $E_{{\theta_2^\prime},6{\theta_0^\prime}}$ by  $\varphi$, respectively:
\begin{align}
\varphi:\quad
\begin{matrix}
z_{20}&z_{21}&z_{22}\\
z_{00}&z_{01}&z_{02}\\
z_{10}&z_{11}&z_{12}\\
\end{matrix}
\quad
\longmapsto
\quad
\begin{matrix}
p_0&p_1&p_2\\
p_3&p_4&p_5\\
p_6&p_7&p_8\\
\end{matrix}\ .
\label{eq:ztop}
\end{align}

\section{Addition formula for the tropical Hesse pencil}
\subsection{Parametrization of the complex torus}
In \cite{KKNT09}, we apply the procedure of ultradiscretization to the level-three theta functions, and obtain piecewise linear functions which parametrize the complement of the tentacles of the tropical Hesse curve.
We recall the result here.

Let $K$ and $\varepsilon$ be positive numbers.
Let us fix $\tau$ as follows
\begin{align}
\tau
=
-\frac{3K}{9K+2\pi i \varepsilon}.
\label{eq:tauud}
\end{align}
For this choice of $\tau$, a point $z\in\C/L_{\tau}$ is written as follows
\begin{align}
z
&=
(-\tau)a+(3\tau+1)b\nn\\
&=
\frac{3Ka}{9K+2\pi i \varepsilon}
+
\frac{2\pi i \varepsilon b}{9K+2\pi i \varepsilon},
\label{eq:zparall}
\end{align}
where $0\leq a,b<1$.
Introducing such a new variable $u\in\R$ that
\begin{align*}
a
=
\frac{u}{3K}
\left(
1+\xi_\varepsilon^2
\right),
\end{align*}
where $\xi_\varepsilon={2\pi\varepsilon}/{9K}$, \eqref{eq:zparall} reduces to
\begin{align}
z
=
\frac{\left(1-i\xi_\varepsilon \right)u}
{9K}
+
\frac{i\xi_\varepsilon b}{1+i\xi_\varepsilon}.
\label{eq:zparallu}
\end{align}
Since $0\leq a<1$, we have
\begin{align}
0\leq u< \frac{3K}{1+\xi_\varepsilon^2}<3K.
\label{eq:urange}
\end{align}

If we take the limit $\varepsilon\to0$ then we have
\begin{align*}
\tau\to-\frac{1}{3},
\qquad
\xi_\varepsilon\to0,
\qquad
\mbox{and}
\qquad
z\to\frac{u}{9K}.
\end{align*}
Hence we obtain
\begin{align*}
\begin{matrix}
z_{20}&z_{21}&z_{22}\\[3pt]
z_{00}&z_{01}&z_{02}\\[3pt]
z_{10}&z_{11}&z_{12}\\[3pt]
\end{matrix}
\quad
\longrightarrow
\quad
\begin{matrix}
0&0&0\\[3pt]
\frac{1}{9}&\frac{1}{9}&\frac{1}{9}\\[3pt]
\frac{2}{9}&\frac{2}{9}&\frac{2}{9}\\[3pt]
\end{matrix}
\qquad
(\varepsilon\to0).
\end{align*}
In terms of the variable $u$, we put the limit of zeros $z_{kj}$ $(k,j=0,1,2)$ as follows
\begin{align}
&u_2:=\lim_{\varepsilon\to0}9Kz_{20}=\lim_{\varepsilon\to0}9Kz_{21}=\lim_{\varepsilon\to0}9Kz_{22}=0
\label{eq:udp2}\\
&u_0:=\lim_{\varepsilon\to0}9Kz_{00}=\lim_{\varepsilon\to0}9Kz_{01}=\lim_{\varepsilon\to0}9Kz_{02}=K
\label{eq:udp0}\\
&u_1:=\lim_{\varepsilon\to0}9Kz_{10}=\lim_{\varepsilon\to0}9Kz_{11}=\lim_{\varepsilon\to0}9Kz_{12}=2K.
\label{eq:udp1}
\end{align}

Let us consider a line $l_\varepsilon$ in $\C$ along with the $a$-axis
\begin{align*}
l_\varepsilon
=
\left\{
\left.
\frac{\left(1-i\xi_\varepsilon \right)u}
{9K}
\ \right|\ 
u\in\R
\right\}.
\end{align*}
Then the circle $l_\varepsilon/\tau\Z$ is contained in the complex torus $\C/L_\tau$.
We define the tropical Jacobian $J(C_K)$ of the tropical Hesse curve $C_K$ as follows
\begin{align*}
\lim_{\varepsilon\to0}l_\varepsilon/\tau\Z
\simeq
\R/3K\Z
=
\{u\in \R\ |\ 0\leq u<3K\}
=:
J(C_K).
\end{align*}

\begin{proposition}
\normalfont
Let $\tau$ be as in \eqref{eq:tauud}.
Then the complex torus $\C/L_{\tau}$ converges into $J(C_K)$ in the limit $\varepsilon\to0$ with respect to the Hausdorff metric.
\end{proposition}

(Proof)\quad
Let the point $z\in\C/L_{\tau}$ be as in \eqref{eq:zparallu}.
Then
\begin{align*}
\inf_{v\in J(C_K)} d\left(9Kz,J(C_K)\right)
&\leq
d(9Kz,u)\\
&=
\left|
{-i u}
+
\frac{9Ki b}{1+i\xi_\varepsilon}
\right|\xi_\varepsilon\\
&<M\varepsilon
\end{align*}
for some $M>0$.
Similarly, for sufficiently small $\varepsilon>0$,  it follows form \eqref{eq:urange} that we have
\begin{align*}
v<\frac{3K}{1+\xi_\varepsilon}
\end{align*}
for any $v\in J(C_K)$, and hence we can take such $z$ that
\begin{align*}
z
=
\frac{\left(1-i\xi_\varepsilon \right)v}
{9K}
+
\frac{i\xi_\varepsilon b}{1+i\xi_\varepsilon}.
\end{align*}
Thus we have
\begin{align*}
\inf_{z\in \C/L_{\tau}} d\left(\C/L_{\tau},v\right)
&\leq
d\left({\left(1-i\xi_\varepsilon \right)v}
+
\frac{9Ki\xi_\varepsilon b}{1+i\xi_\varepsilon},v\right)\\
&=
\left|
{-i v}
+
\frac{9K i b}{1+i\xi_\varepsilon}
\right|\xi_\varepsilon\\
&<M^\prime\varepsilon.
\end{align*}
for some $M^\prime>0$.
This completes the proof.
\qed

\subsection{Ultradiscretization}
Now we show that the points on the $a$-axis in the complex torus $\C/L_\tau$ correspond to  that on the real part of the Hesse cubic curve.

\begin{proposition}\label{prop:real}
\normalfont
Let $\tau$ be as in \eqref{eq:tauud}.
Then $\varphi$ maps the points on the circle $l_\varepsilon/\tau\Z$ into $E_{{\theta_2^\prime},6{\theta_0^\prime}}\cap\P^2(\R)$, the real part of the Hesse cubic curve.
\end{proposition}

(Proof)\quad
By using the formula concerning the modular transformation of the level-three theta functions (see proposition 4.3 in \cite{KKNT09}), we have
\begin{align*}
&\theta_{k}(z,\tau)
=
e^{\frac{-9\pi iz^{2}}{3\tau+1}}(3\tau+1)^{-\frac{1}{2}}e^{\frac{\pi i}{4}}
\vartheta_{(\frac{k}{3}-\frac{7}{6},\frac{3}{2})}\left(
\frac{3z}{3\tau+1},\frac{3\tau}{3\tau+1}
\right).
\end{align*}
Since $z$ is assumed to be on $l_\varepsilon/\tau\Z$, we can put $z$ be as in \eqref{eq:zparallu} with $b=0$.
Then we obtain
\begin{align}
&\vartheta_{(\frac{k}{3}-\frac{7}{6},\frac{3}{2})}\left(
\frac{3z}{3\tau+1},\frac{3\tau}{3\tau+1}
\right)
=
(-1)^{k}i
e^{\frac{u^{2}}{2K\varepsilon}}\nn\\
&\qquad\times
\sum_{n\in\Z}
\exp
\left[
\left(
n+\frac{k}{3}-\frac{7}{6}
\right)
\frac{3\xi_\varepsilon^2}{\varepsilon} u
-\frac{9K}{2\varepsilon}
\left(
\frac{u-(k+1)K}{3K}
-n+\frac{3}{2}
\right)^{2}
\right]
(-1)^{n}.
\label{eq:thetareal}
\end{align}
The imaginary part of the functions $\theta_0(z,\tau)$, $\theta_1(z,\tau)$, and $\theta_2(z,\tau)$ appear only in the following common factor 
\begin{align*}
e^{\frac{-9\pi iz^{2}}{3\tau+1}}(3\tau+1)^{-\frac{1}{2}}e^{\frac{\pi i}{4}}i.
\end{align*}
Therefore, we have $\varphi(z)\in E_{{\theta_2^\prime},6{\theta_0^\prime}}\cap\P^2(\R)$.
\qed

There exist three zeros $z_{20}$, $z_{00}$, and $z_{10}$ of the level-three theta functions on $l_\varepsilon/\tau\Z$ (see figure \ref{fig:lattice}).
These zeros divide $l_\varepsilon/\tau\Z$ into three open intervals denoted by $d_1$, $d_2$, and $d_3$:
\begin{align*}
d_j
:=&
\left\{
-\tau a\in\C/L_\tau\ \left|\ \frac{j-1}{3}< a<\frac{j}{3}\right.
\right\}\\
=&
\left\{
\frac{\left(1-i\xi_\varepsilon \right)}{9K}u\in\C/L_\tau\ \left|\ \frac{(j-1)K}{1+\xi_\varepsilon^2}< u<\frac{jK}{1+\xi_\varepsilon^2}\right.
\right\}
\qquad
(j=1,2,3).
\end{align*}
Noticing \eqref{eq:ztop}, we have
\begin{align*}
\varphi(z_{20})=p_{0}=(\infty,\infty),\quad
\varphi(z_{00})=p_{3}=(0,-1),\quad
\varphi(z_{10})=p_{6}=(-1,0)
\end{align*}
in the inhomogeneous coordinate $(x:=x_1/x_0,y:=x_2/x_0)$ of $\P^2(\C)$, and hence we obtain the following (see figure \ref{fig:HesseCurve})
\begin{align}
&\varphi(d_1)
=
\left\{
\left.(x,y)\in E_{{\theta_2^\prime},6{\theta_0^\prime}}\cap\P^2(\R)\ \right|\ x>0,\ y<0
\right\}\label{eq:phid0}\\
&\varphi(d_2)
=
\left\{
\left.(x,y)\in E_{{\theta_2^\prime},6{\theta_0^\prime}}\cap\P^2(\R)\ \right|\ x<0,\ y<0
\right\}\label{eq:phid1}\\
&\varphi(d_3)
=
\left\{
\left.(x,y)\in E_{{\theta_2^\prime},6{\theta_0^\prime}}\cap\P^2(\R)\ \right|\ x<0,\ y>0
\right\}.\label{eq:phid2}
\end{align}

We define the open subsets ${D}_1$, ${D}_2$, and ${D}_3$ of $J(C_K)$ as follows
\begin{align*}
D_j
:=
\lim_{\varepsilon\to0}d_j
=
\left\{
u\in J(C_K)\ |\ {(j-1)K}< u<{jK}
\right\}
\qquad
(j=1,2,3).
\end{align*}
Then we have $J(C_K)=\bigcup_{j=0}^2 \left(D_{j+1}\cup u_j\right)$, where $u_k\equiv(k+1) K$ (mod 3) is the limiting point of the zeros of $\theta_k(z,\tau)$ for $k=0,1,2$ (see (\ref{eq:udp2}-\ref{eq:udp1})).

Next we consider the amoeba of the real part of $E_{{\theta_2^\prime},6{\theta_0^\prime}}$ which is defined as the set of points $(\varepsilon\log |x|,\varepsilon\log |y|)$ satisfying $(x,y)\in E_{{\theta_2^\prime},6{\theta_0^\prime}}\cap\P^2(\R)$.
Let $z$ be a point on the open set $d_{1}\cup d_{2}\cup d_{3}$.
Then we have $\theta_{k}(z)\neq0$ for $k=0,1,2$.
Since $(x,y)=\varphi(z)=(\theta_{0}(z)/\theta_{2}(z),\theta_{1}(z)/\theta_{2}(z))$, we have (see \eqref{eq:thetareal})
\begin{align*}
\varepsilon\log |x|
&=
\varepsilon\log\left|\frac{\theta_0(z)}{\theta_2(z)}\right|\\
&=
\varepsilon\log
\left|
\frac
{\DIS\sum_{n\in\Z}
\exp
\left[
\left(
n-\frac{7}{6}
\right)
\frac{3\xi_\varepsilon^2}{\varepsilon} u
-\frac{9K}{2\varepsilon}
\left(
\frac{u-K}{3K}
-n+\frac{3}{2}
\right)^{2}
\right]
(-1)^{n}}
{\DIS\sum_{n\in\Z}
\exp
\left[
\left(
n-\frac{1}{2}
\right)
\frac{3\xi_\varepsilon^2}{\varepsilon} u
-\frac{9K}{2\varepsilon}
\left(
\frac{u-3K}{3K}
-n+\frac{3}{2}
\right)^{2}
\right]
(-1)^{n}}
\right|
\end{align*}
and 
\begin{align*}
\varepsilon\log |y|
&=
\varepsilon\log\left|\frac{\theta_1(z)}{\theta_2(z)}\right|\\
&=
\varepsilon\log
\left|
\frac
{\DIS\sum_{n\in\Z}
\exp
\left[
\left(
n-\frac{5}{6}
\right)
\frac{3\xi_\varepsilon^2}{\varepsilon} u
-\frac{9K}{2\varepsilon}
\left(
\frac{u-2K}{3K}
-n+\frac{3}{2}
\right)^{2}
\right]
(-1)^{n}}
{\DIS\sum_{n\in\Z}
\exp
\left[
\left(
n-\frac{1}{2}
\right)
\frac{3\xi_\varepsilon^2}{\varepsilon} u
-\frac{9K}{2\varepsilon}
\left(
\frac{u-3K}{3K}
-n+\frac{3}{2}
\right)^{2}
\right]
(-1)^{n}}
\right|.
\end{align*}

Define the piecewise linear functions
\begin{align*}
&\tilde c(u)
:=
-\frac{9K}{2}
\left\{
\left(\left(
\frac{u-K}{3K}-\frac{1}{2}
\right)\right)
\right\}^2
+
\frac{9K}{2}
\left\{
\left(\left(
\frac{u-3K}{3K}-\frac{1}{2}
\right)\right)
\right\}^2\\
&\tilde s(u)
:=
-\frac{9K}{2}
\left\{
\left(\left(
\frac{u-2K}{3K}-\frac{1}{2}
\right)\right)
\right\}^2
+
\frac{9K}{2}
\left\{
\left(\left(
\frac{u-3K}{3K}-\frac{1}{2}
\right)\right)
\right\}^2,
\end{align*}
where the function $\left(\left(\ \right)\right):\R\to[0,1)$ is defined as follows
\begin{align*}
\left(\left(u\right)\right)
=
u-{\rm Floor}(u).
\end{align*}
Also define the subset $R\subset\R$
\begin{align*}
R
:=
\bigcup_{j=1}^3
\left\{
a\in\R\ \left|\ 
\frac{j-1}{3}< a< \frac{j}{3}\right.
\right\}.
\end{align*}
Then we obtain the following proposition.
\begin{proposition}\normalfont
\label{prop:realconv}
Let $\tau$ be as in \eqref{eq:tauud}.
Assume $a\in R$.
Then the functions 
\begin{align}
\varepsilon\log\left|\frac{\theta_0(-a\tau)}{\theta_2(-a\tau)}\right|
\qquad
\mbox{and}
\qquad
\varepsilon\log\left|\frac{\theta_1(-a\tau)}{\theta_2(-a\tau)}\right|
\label{eq:tocs}
\end{align}
uniformly converge into 
\begin{align}
\tilde c(3Ka)
\qquad
\mbox{and}
\qquad
\tilde s(3Ka)
\label{eq:tildecs}
\end{align}
in the limit $\varepsilon\to0$ in the wider sense, respectively.
\end{proposition}

(Proof)\quad
By proposition \ref{prop:real}, $\varphi(z)$ is a point on the real part of $E_{{\theta_2^\prime},6{\theta_0^\prime}}$ for $z\in\{-a\tau\ |\ 0\leq a<1\}$.
Let $\delta$ be an arbitrary positive number less than 1/6.
Then we can take $a$ satisfying $j/3+\delta\leq a\leq (j+1)/3-\delta$ for any $j\in\{0,1,2\}$.
For this choice of $a$, we have $-a\tau\in d_j$, and hence $\theta_k(-a\tau)$ ($k=0,1,2$) does not vanish.
Thus we can define the functions \eqref{eq:tocs} on the compact set 
\begin{align*}
R_\delta
:=
\bigcup_{j=1}^3
\left\{
a\in\R\ \left|\ 
\frac{j-1}{3}+\delta\leq a\leq \frac{j}{3}-\delta\right.
\right\}.
\end{align*}
If we take $a$ as above, then we have $jK+3K\delta\leq 3Ka\leq (j+1)K-3K\delta$, and hence $3Ka$ is contained in $D_j$ because of the assumption $0<\delta<1/6$.
Thus the functions \eqref{eq:tildecs} can also be defined on $R_{\delta}$. 

We can estimate $\theta_0(-a\tau)$ as follows.
Put 
\begin{align*}
m=n-n_0(u),
\qquad
n_0(u)={\rm Floor}\left(\frac{u-K}{3K}\right)+2.
\end{align*}
Then we have
\begin{align*}
&\sum_{n\in\Z}
\exp
\left[
\left(
n-\frac{7}{6}
\right)
\frac{3\xi_\varepsilon^2}{\varepsilon} u
\right]
\exp
\left[
-\frac{9K}{2\varepsilon}
\left(
\frac{u-K}{3K}
-n+\frac{3}{2}
\right)^{2}
\right](-1)^n\\
&
\quad=
\sum_{m\in\Z}
\exp
\left[
\left(
m+n_0(u)-\frac{7}{6}
\right)
\frac{3\xi_\varepsilon^2}{\varepsilon} u
\right]
\exp
\left[
-\frac{9K}{2\varepsilon}
\left\{
\left(\left(
\frac{u-K}{3K}
\right)\right)
-m-\frac{1}{2}
\right\}^{2}
\right](-1)^{m+n_0(u)}\\
&\quad
=
\exp
\left[
-\frac{9K}{2\varepsilon}
\left\{
\left(\left(
\frac{u-K}{3K}
\right)\right)-\frac{1}{2}
\right\}^{2}
+
\left(
n_0(u)-\frac{7}{6}
\right)
\frac{3\xi_\varepsilon^2}{\varepsilon} u
\right](-1)^{n_0(u)}\\
&\qquad\qquad
\times
\sum_{m\in\Z}
\exp
\left[
-\frac{9K}{2\varepsilon}
\left\{
m+1
-2\left(\left(
\frac{u-K}{3K}
\right)\right)
-
\frac{2\xi_\varepsilon^2}{3K} u
\right\}m
\right](-1)^m.
\end{align*}
Noting $u\in D_1\cup D_2\cup D_3$, for sufficiently small $\varepsilon$, we have
\begin{align*}
&-\frac{9K}{2\varepsilon}
\left\{
m+1
-2\left(\left(
\frac{u-K}{3K}
\right)\right)
-
\frac{2\xi_\varepsilon^2}{3K} u
\right\}
<0
\qquad
\mbox{for $m>>0$}\\
&-\frac{9K}{2\varepsilon}
\left\{
m+1
-2\left(\left(
\frac{u-K}{3K}
\right)\right)
-
\frac{2\xi_\varepsilon^2}{3K} u
\right\}
>0
\qquad
\mbox{for $m<<0$}.
\end{align*}
Hence, for any $u\in D_1\cup D_2\cup D_3$,  there exists such $0<r<1$ that
\begin{align*}
\exp
\left[
-\frac{9K}{2\varepsilon}
\left\{
m+1
-2\left(\left(
\frac{u-K}{3K}
\right)\right)
-
\frac{2\xi_\varepsilon^2}{3K} u
\right\}m
\right]
<
r^{|m|}
\qquad
\mbox{except for finite $m\in\Z$}.
\end{align*}
Therefore, we have
\begin{align*}
\varepsilon\log\left|{\theta_0(-a\tau)}\right|
=
-\frac{9K}{2}
\left\{
\left(\left(
\frac{u-K}{3K}
\right)\right)-\frac{1}{2}
\right\}^{2}
+
o(1).
\end{align*}
Similarly, for $\theta_2(-a\tau)$, we have
\begin{align*}
\varepsilon\log\left|{\theta_2(-a\tau)}\right|
=
-\frac{9K}{2}
\left\{
\left(\left(
\frac{u-3K}{3K}
\right)\right)-\frac{1}{2}
\right\}^{2}
+
o(1).
\end{align*}
Thus the function $\varepsilon\log\left|{\theta_0(-a\tau)}/{\theta_2(-a\tau)}\right|$ uniformly converges into $\tilde c(3Ka)$ in the limit $\varepsilon\to0$ on $R_\delta$.
The case for $\tilde s(3Ka)$ is similarly shown \cite{KKNT09}. 
\qed

The piecewise linear functions $\tilde c$ and $\tilde s$ are defined on $\bigcup_{j=1}^3 D_j=J(C_K)\setminus\{u_0,u_1,u_2\}$.
If we extend them to be continuous functions on $J(C_K)$ then their values on the points $u_2,u_0,u_1$ are uniquely determined as follows
\begin{align}
&
\tilde c(u_2)=K
&&
\tilde c(u_0)=-K
&&
\tilde c(u_1)=0
&\nn\\
&
\tilde s(u_2)=K
&&
\tilde s(u_0)=0
&&
\tilde s(u_1)=-K.
&\label{eq:limofu}
\end{align}
The extended continuous piecewise linear functions are also denoted by $\tilde c$ and $\tilde s$ (see figures \ref{fig:xcomp} and \ref{fig:ycomp}).
Note that the points $u_2,u_0,u_1$ are mapped into the vertices of $\bar C_K$:
\begin{align*}
\left(\tilde c(u_2),\tilde s(u_2)\right)=V_1,
\qquad
\left(\tilde c(u_0),\tilde s(u_0)\right)=V_2,
\qquad
\left(\tilde c(u_1),\tilde s(u_1)\right)=V_3.
\end{align*}
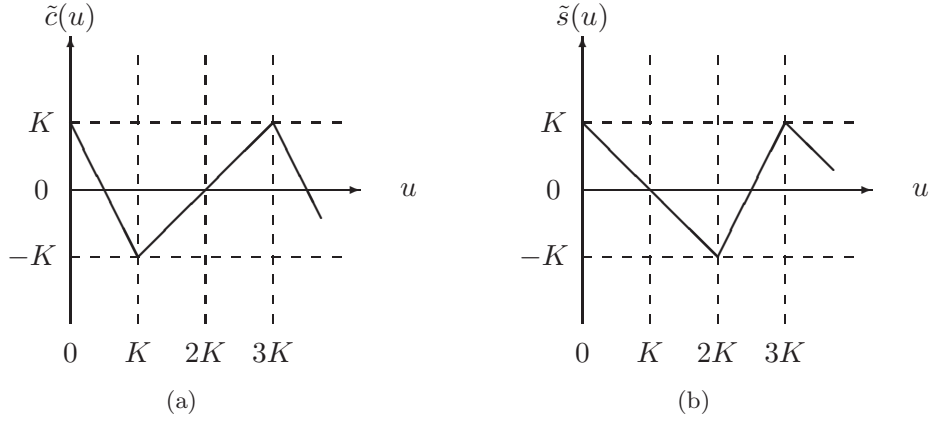
\begin{figure}[htbp]
\centering
{\unitlength=.05in{\def\arraystretch{1.0}
\subfigure[]{
\begin{picture}(50,40)(-43,-20)
\put(-30,-1){\vector(1,0){30}}
\put(-30,-15){\vector(0,1){30}}
\put(5,-1){\makebox(0,0){$u$}}
\put(-30,17){\makebox(0,0){$\tilde c(u)$}}
\dashline[10]{1}(-23,-15)(-23,13)
\dashline[10]{1}(-9,-15)(-9,13)
\dashline[10]{1}(-16,-15)(-16,13)
\dashline[10]{1}(-30,-8)(-2,-8)
\dashline[10]{1}(-30,6)(-2,6)
\thicklines
\put(-30,6){\line(1,-2){7}}
\put(-23,-8){\line(1,1){14}}
\put(-9,6){\line(1,-2){5}}
\put(-30,-18){\makebox(0,0){$0$}}
\put(-23,-18){\makebox(0,0){$K$}}
\put(-16,-18){\makebox(0,0){$2K$}}
\put(-9,-18){\makebox(0,0){$3K$}}
\put(-33,-1){\makebox(0,0){$0$}}
\put(-34,-8){\makebox(0,0){$-K$}}
\put(-33,6){\makebox(0,0){$K$}}
\end{picture}
\label{fig:xcomp}}
\subfigure[]{
\begin{picture}(50,40)(-43,-20)
\put(-30,-1){\vector(1,0){30}}
\put(-30,-15){\vector(0,1){30}}
\put(5,-1){\makebox(0,0){$u$}}
\put(-30,17){\makebox(0,0){$\tilde s(u)$}}
\dashline[10]{1}(-23,-15)(-23,13)
\dashline[10]{1}(-9,-15)(-9,13)
\dashline[10]{1}(-16,-15)(-16,13)
\dashline[10]{1}(-30,-8)(-2,-8)
\dashline[10]{1}(-30,6)(-2,6)
\thicklines
\put(-30,6){\line(1,-1){14}}
\put(-16,-8){\line(1,2){7}}
\put(-9,6){\line(1,-1){5}}
\put(-30,-18){\makebox(0,0){$0$}}
\put(-23,-18){\makebox(0,0){$K$}}
\put(-16,-18){\makebox(0,0){$2K$}}
\put(-9,-18){\makebox(0,0){$3K$}}
\put(-33,-1){\makebox(0,0){$0$}}
\put(-34,-8){\makebox(0,0){$-K$}}
\put(-33,6){\makebox(0,0){$K$}}
\end{picture}
\label{fig:ycomp}}
}}
\caption{(a) $\tilde c:J(C_K)\to\R$．
(b) $\tilde s:J(C_K)\to\R$．}
\end{figure}

Let us introduce a map $\tilde\varphi: J(C_K)\to \R^2\subset\P^{2,trop}$,
\begin{align}
\tilde\varphi:\quad
u\longmapsto(\tilde c(u),\tilde s(u)).
\label{eq:tildephi}
\end{align}
This map induces an isomorphism $J(C_K)\simeq \bar C_K$. 
Therefore, we have
\begin{align*}
\left\{
(\tilde c(u),\tilde s(u))\in\R^2\ |\ 
u\in J(C_K)
\right\}
=
\bar C_K.
\end{align*}
Thus we obtain the piecewise linear functions $\tilde c(u)$ and $\tilde s(u)$ which parametrize the tropical Hesse pencil.
Since $\tilde\varphi(0)=(\tilde c(0),\tilde s(0))=V_1$, the additive group structure of $(\bar C_K,V_1)$, equipped with the unit of addition $V_1$, is induced from that of $J(C_K)=\R/3K\Z$ via the group isomorphism $\tilde\varphi$.
We denote the addition on the tropical Hesse curve by $\uplus:\bar C_K\times \bar C_K\to \bar C_K$.

It is easy to see that we have
\begin{align}
&\tilde\varphi(D_1)
=
\left\{
(X,Y)\in \bar C_K\subset\P^{2,trop}\ |\ Y>0,\ Y>X
\right\}
=
E_1^\circ
\label{eq:tildephid0}\\
&\tilde\varphi(D_2)
=
\left\{
(X,Y)\in \bar C_K\subset\P^{2,trop}\ |\ X<0,\ Y<0
\right\}
=
E_2^\circ
\label{eq:tildephid1}\\
&\tilde\varphi(D_3)
=
\left\{
(X,Y)\in \bar C_K\subset\P^{2,trop}\ |\ X>0,\ Y<X
\right\}
=
E_3^\circ,
\label{eq:tildephid2}
\end{align}
where $E_j^\circ:=E_j\setminus\{V_j,V_{j+1}\}$ stands for the interior of $E_j$.

Let ${\rm Log}:\R^2\to\R^2$ be the map
\begin{align*}
{\rm Log}\mkern2mu : (x,y)\mapsto (\varepsilon\log\left|x\right|,\varepsilon\log\left|y\right|).
\end{align*}
Let the amoeba of the real part of $E_{{\theta_2^\prime},6{\theta_0^\prime}}$ be
\begin{align*}
A_\varepsilon
:=
\left\{
\left({\rm Log}\circ\varphi\right)(z)\ \left|\ 
z\in \bigcup_{j=1}^3d_j\right.
\right\}
\end{align*}
It follows from proposition \ref{prop:realconv} that we have the commutative diagram
\begin{align*}
\begin{CD}
l_\varepsilon/\tau\Z\setminus\{z_{20},z_{00},z_{10}\} 
@ > \varepsilon\to0 >> 
\R/3K\Z\setminus\{u_{2},u_{0},u_{1}\} \\
@ V {\rm Log}\mkern2mu\circ\varphi VV
@ VV \tilde\varphi V\\
A_\varepsilon
@ > \varepsilon\to0 >> 
\bar C_K\setminus\{V_{1},V_{2},V_{3}\} .
\end{CD}
\end{align*}

\subsection{Tropical Hesse configuration}
Now we consider the tropical counterpart of the Hesse configuration.
Remember that the Hesse configuration consists of the 9 inflection points $p_{0},p_{1},\cdots,p_{8}$ and the 12 inflections lines, which compose the singular members $E_{0,1}$, $E_{1,-3}$, $E_{1,-3\zeta_{3}}$, and $E_{1,-3\zeta_{3}^{2}}$ of the pencil (see table \ref{tab:Hesseconfig}).

Fix $\tau$ as in \eqref{eq:tauud}.
We consider a map $\eta:E_{{\theta_2^\prime},6{\theta_0^\prime}}\to \bar C_K$ so defined that the diagram commute:
\begin{align*}
\begin{CD}
\C/L_\tau
@ > \varepsilon\to0 >> 
J(C_K)\\
@ V \varphi VV
@ VV \tilde\varphi V\\
E_{{\theta_2^\prime},6{\theta_0^\prime}}
@ > \eta >>
\bar C_K.\\
\end{CD}
\end{align*}
The inflection points of $E_{{\theta_2^\prime},6{\theta_0^\prime}}$ are mapped into the vertices of $\bar C_K$ by $\eta$ as follows
\begin{align*}
&\eta:\ 
p_0,\ p_1,\ p_2
\overset{\varphi^{-1}}{\longmapsto}
z_{20},\ z_{21},\ z_{22}
\overset{\varepsilon\to0}{\longrightarrow}
u_2
\overset{\tilde\varphi}{\longmapsto}
V_1\\
&\eta:\ 
p_3,\ p_4,\ p_5
\overset{\varphi^{-1}}{\longmapsto}
z_{00},\ z_{01},\ z_{02}
\overset{\varepsilon\to0}{\longrightarrow}
u_0
\overset{\tilde\varphi}{\longmapsto}
V_2\\
&\eta:\ 
p_6,\ p_7,\ p_8
\overset{\varphi^{-1}}{\longmapsto}
z_{10},\ z_{11},\ z_{12}
\overset{\varepsilon\to0}{\longrightarrow}
u_1
\overset{\tilde\varphi}{\longmapsto}
V_3.
\end{align*}
Thus the tropical counterparts of the Hesse configuration consists of the vertices of $\bar C_K$ and the lines passing through them.
Moreover, the lines passing through the vertices should compose the singular members of the tropical Hesse pencil.

Table \ref{tab:udsingcurve} shows that there exist two singular members $C_0$ and $C_\infty$ in the tropical Hesse pencil.
The member $C_0$ is a triple tropical line defined by \eqref{eq:defpolysing2}.
Each of the three points $V_1$, $V_2$, and $V_3$ is clearly on each of the three tentacles of $C_0$.
On the other hand, the singular member $C_\infty$ is the boundary of $\P^{2,trop}$ defined by \eqref{eq:defpolysing1} (see \eqref{eq:cinf}).
Since the points $V_1$, $V_2$, and $V_3$ are contained inside of $\P^{2,trop}$, it looks that they are not on $C_\infty$.
However, noticing the linear equivalence relation $\sim$, which identifies all points on a tentacle, and the fact a tentacle to intersect a boundary of $\P^{2,trop}$, we can conclude that all the points  $V_1$, $V_2$, and $V_3$ are contained in $C_\infty$.
Thus the tropical counterpart of the Hesse configuration consists of three points and four lines which satisfy the following two conditions;
\begin{itemize}
\item each line passes through at least one of the three points and 
\item each point lies on two of the four lines.
\end{itemize}
We illustrate the Hesse configuration and its tropical counterpart in table \ref{tab:tropHesseconfig}.

\begin{table}[htbp]
\renewcommand{\arraystretch}{1.2}
\begin{center}
\caption{The correspondence between the Hesse configuration and its tropical counterpart.}
\begin{tabular}{clrc}\label{tab:tropHesseconfig}
\\\HLINE
\multicolumn{2}{c}{Hesse pencil}&
\multicolumn{2}{c}{Tropical Hesse pencil}\\
Singular curves&Hesse configuration&Hesse configuration&Singular curves\\\hline
$E_{0,1}$
&
{\unitlength=.05in{\def\arraystretch{1.0}
\begin{picture}(13,7)(0,3)
\thicklines
\put(-2,0){\line(1,0){18}}
\put(-1,-1){\line(1,1){9}}
\put(15,-1){\line(-1,1){9}}
\put(3,0){\circle*{1}}
\put(7,0){\circle*{1}}
\put(11,0){\circle*{1}}
\put(1,1){\circle*{1}}
\put(3.5,3.5){\circle*{1}}
\put(6,6){\circle*{1}}
\put(13,1){\circle*{1}}
\put(10.5,3.5){\circle*{1}}
\put(8,6){\circle*{1}}
\end{picture}
}}
&
{\unitlength=.05in{\def\arraystretch{1.0}
\begin{picture}(13,7)(0,4)
\thicklines
\put(0,0){\line(1,0){9}}
\put(0,0){\line(0,1){9}}
\put(9,0){\line(-1,1){9}}
\put(4.5,0){\circle*{1}}
\put(0,4.5){\circle*{1}}
\put(4.5,4.5){\circle*{1}}
\end{picture}
}}
&
$C_\infty$
\\[15pt]\hline
$E_{1,-3}$
&
{\unitlength=.05in{\def\arraystretch{1.0}
\begin{picture}(13,7)(0,3)
\thicklines
\put(-2,0){\line(1,0){18}}
\put(-1,-1){\line(1,1){9}}
\put(15,-1){\line(-1,1){9}}
\put(3,0){\circle*{1}}
\put(7,0){\circle*{1}}
\put(11,0){\circle*{1}}
\put(1,1){\circle*{1}}
\put(3.5,3.5){\circle*{1}}
\put(6,6){\circle*{1}}
\put(13,1){\circle*{1}}
\put(10.5,3.5){\circle*{1}}
\put(8,6){\circle*{1}}
\end{picture}
}}
&&\\[15pt]
$E_{1,-3\zeta_3}$
&
{\unitlength=.05in{\def\arraystretch{1.0}
\begin{picture}(13,7)(0,3)
\thicklines
\put(-2,0){\line(1,0){18}}
\put(-1,-1){\line(1,1){9}}
\put(15,-1){\line(-1,1){9}}
\put(3,0){\circle*{1}}
\put(7,0){\circle*{1}}
\put(11,0){\circle*{1}}
\put(1,1){\circle*{1}}
\put(3.5,3.5){\circle*{1}}
\put(6,6){\circle*{1}}
\put(13,1){\circle*{1}}
\put(10.5,3.5){\circle*{1}}
\put(8,6){\circle*{1}}
\end{picture}
}}
&
{\unitlength=.05in{\def\arraystretch{1.0}
\begin{picture}(13,7)(0,4)
\thicklines
\put(4.5,4.5){\line(-1,0){5}}
\put(4.5,4.5){\line(0,-1){5}}
\put(4.5,4.5){\line(1,1){5}}
\put(4.5,2){\circle*{1}}
\put(2,4.5){\circle*{1}}
\put(6,6){\circle*{1}}
\end{picture}
}}
&
$C_0$\\[15pt]
$E_{1,-3\zeta_3^2}$
&
{\unitlength=.05in{\def\arraystretch{1.0}
\begin{picture}(13,7)(0,3)
\thicklines
\put(-2,0){\line(1,0){18}}
\put(-1,-1){\line(1,1){9}}
\put(15,-1){\line(-1,1){9}}
\put(3,0){\circle*{1}}
\put(7,0){\circle*{1}}
\put(11,0){\circle*{1}}
\put(1,1){\circle*{1}}
\put(3.5,3.5){\circle*{1}}
\put(6,6){\circle*{1}}
\put(13,1){\circle*{1}}
\put(10.5,3.5){\circle*{1}}
\put(8,6){\circle*{1}}
\end{picture}
}}
&&\\[15pt]
\HLINE
\end{tabular}
\end{center}
\end{table}

\subsection{Ultradiscrete elliptic functions}
Now we construct the addition formula for the points on the tropical Hesse curve via the ultradiscretization of that for the Hesse cubic curve\footnote{In \cite{KKNT09}, we have already presented the duplication formula for the tropical Hesse curve.}.
For this purpose, we introduce elliptic functions defined by the ratios of the level-three theta functions:
\begin{align*}
c(z)
:=
\frac{\theta_{0}(z,\tau)}{\theta_{2}(z,\tau)}
\qquad
s(z)
:=
\frac{\theta_{1}(z,\tau)}{\theta_{2}(z,\tau)}.
\end{align*}
It can be easily checked that the following holds
\begin{align*}
&c(z+1)
=
c(z)&
&c(z+\tau)
=
c(z)\\
&s(z+1)
=
s(z)&
&s(z+\tau)
=
s(z).
\end{align*}
Therefore $c(z)$ and $s(z)$ are elliptic functions which have the double periodicity with respect to the translations $z\to z+1$ and $z\to z+\tau$.

In proposition \ref{prop:realconv}, we show that the following holds for any $z=-a\tau\in d_j$ ($j=1,2,3$)
\begin{align*}
\lim_{\varepsilon\to0}\varepsilon\log c(z)
=
\tilde c(u)
\qquad
\mbox{and}
\qquad
\lim_{\varepsilon\to0}\varepsilon\log s(z)
=
\tilde s(u),
\end{align*}
where $u=3Ka\in D_j$ and $\tau$ is assumed to be as in \eqref{eq:tauud}.
Therefore, we call $\tilde c$ and $\tilde s$ ultradiscrete elliptic functions.
Note that $\tilde c(u)$ and $\tilde s(u)$ have single periodicity with respect to the translation $u\to u+3K$ (see figures \ref{fig:xcomp} and \ref{fig:ycomp}).

The addition formulae for the elliptic functions $c(z)$ and $s(z)$ immediately follow from that for the level-three theta functions \eqref{eq:addtheta1}, \eqref{eq:addtheta2}, and \eqref{eq:addtheta3}:
\begin{subequations}
\begin{align}
&c(z+w)
=
\frac{s(z)-c(z)^{2}c(w)s(w)}{c(z)c(w)^{2}-s(z)^{2}s(w)}
\label{eq:addelliptic1a}\\
&s(z+w)
=
\frac{c(z)s(z)s(w)^{2}-c(w)}{c(z)c(w)^{2}-s(z)^{2}s(w)}
\label{eq:addelliptic1b}
\end{align}
\end{subequations}
\begin{subequations}
\begin{align}
&c(z+w)
=
\frac{c(z)s(z)c(w)^{2}-s(w)}{s(z)s(w)^{2}-c(z)^{2}c(w)}
\label{eq:addelliptic2a}\\
&s(z+w)
=
\frac{c(z)-s(z)^{2}c(w)s(w)}{s(z)s(w)^{2}-c(z)^{2}c(w)}
\label{eq:addelliptic2b}
\end{align}
\end{subequations}
\begin{subequations}
\begin{align}
&c(z+w)
=
\frac{c(z)s(w)^{2}-s(z)^{2}c(w)}{c(z)s(z)-c(w)s(w)}
\label{eq:addelliptic3a}\\
&s(z+w)
=
\frac{s(z)c(w)^{2}-c(z)^{2}s(w)}{c(z)s(z)-c(w)s(w)}
\label{eq:addelliptic3b}.
\end{align}
\end{subequations}

\subsection{Addition formula}
Fix $\tau$ as in \eqref{eq:tauud}.
Assume $z$ and $w$ to be points on $l_\varepsilon/\tau\Z$; then we can put them as follows
\begin{align*}
z
=
\frac{\left(1-i\xi_\varepsilon \right)u}
{9K}
\qquad
\mbox{and}
\qquad
w
=
\frac{\left(1-i\xi_\varepsilon \right)v}
{9K},
\end{align*}
where $u,v\in J(C_K)$.

Let us consider \eqref{eq:addelliptic1a}.
By proposition \ref{prop:real}, the elliptic functions $c$ and $s$ are real valued for this choice of $\tau$ and $z,w$.
At first, assume $z,w\in d_1$.
Note that the following holds (see \eqref{eq:phid0})
\begin{align*}
c(z),c(w)>0,\qquad
s(z),s(w)<0.
\end{align*}
Then we have
\begin{align*}
&s(z)
=
-\left|s(z)\right|&&
-c(z)^2c(w)s(w)
=
\left|c(z)^2c(w)s(w)\right|\\
&c(z)c(w)^2
=
\left|c(z)c(w)^2\right|&&
-s(z)^2s(w)
=
\left|s(z)^2s(w)\right|.
\end{align*}
It follows that the denominator of the right hand side of \eqref{eq:addelliptic1a} is always positive, while the sign of the numerator is indeterminate, i.e., it depends on the values of $z$ and $w$.
The left hand side of \eqref{eq:addelliptic1a} has the same sign as the numerator of the right hand side.
Thus we obtain the subtraction-free form of \eqref{eq:addelliptic1a}
\begin{align*}
\left({\left|c(z)c(w)^2\right|+\left|s(z)^2s(w)\right|}\right)
\left|c(z+w)\right|
+
\left|s(z)\right|
=
{\left|c(z)^2c(w)s(w)\right|}
\end{align*}
 if $\left|c(z)^2c(w)s(w)\right|>\left|s(z)\right|$ or 
\begin{align*}
\left({\left|c(z)c(w)^2\right|+\left|s(z)^2s(w)\right|}\right)
\left|c(z+w)\right|
+
{\left|c(z)^2c(w)s(w)\right|}
=
\left|s(z)\right|
\end{align*}
if $\left|c(z)^2c(w)s(w)\right|<\left|s(z)\right|$. 
Therefore, by proposition \ref{prop:realconv}, we obtain
\begin{align}
\tilde c(u+v)
=
\max\left(
\tilde s(u), 2\tilde c(u)+\tilde c(v)+\tilde s(v)
\right)
-
\max\left(
\tilde c(u)+2\tilde c(v),2\tilde s(u)+\tilde s(v)
\right)
\label{eq:elimcsud}
\end{align}
except for $u,v$ satisfying $u+v=K$\footnote{These $u,v$ correspond to $z,w$ such that $c(z+w)=s(z)-c(z)^{2}c(w)s(w)=0$} in the limit $\varepsilon\to0$. 
Since the both hand sides of \eqref{eq:elimcsud} are continuous functions, \eqref{eq:elimcsud} holds even for $u,v$ satisfying $u+v=K$.
Noting \eqref{eq:tildephid0}, we see that \eqref{eq:elimcsud} holds for $u,v$ such that both $(\tilde c(u),\tilde s(u))$ and $(\tilde c(v),\tilde s(v))$ are in $E_1^\circ$, or equivalently, for $u,v\in D_1$.

Next, assume $z\in d_1$ and $w\in d_2$.  
Then we have (see \eqref{eq:phid0} and \eqref{eq:phid1})
\begin{align*}
c(z)>0,\qquad
c(w),s(z),s(w)<0.
\end{align*}
The denominator of the right hand side of \eqref{eq:addelliptic1a} is always positive and the numerator is always negative.
The left hand side of \eqref{eq:addelliptic1a} has the negative sign as well.
Thus we obtain the subtraction-free form
\begin{align*}
\left({\left|c(z)c(w)^2\right|+\left|s(z)^2s(w)\right|}\right)
\left|c(z+w)\right|
=
\left|s(z)\right|
+
{\left|c(z)^2c(w)s(w)\right|}.
\end{align*}
Taking the limit $\varepsilon\to0$, we obtain \eqref{eq:elimcsud} which holds for $u,v$ such that $(\tilde c(u),\tilde s(u))\in {E}_1^\circ$ and $(\tilde c(v),\tilde s(v))\in E_2^\circ$, or equivalently, for $u\in D_1$ and $v\in D_2$.

Thus we observe that \eqref{eq:elimcsud} is the candidate of the addition formula for the tropical Hesse curve.
However, if we assume $z\in d_1$ and $w\in d_3$ then \eqref{eq:elimcsud} does not hold.
Actually, we have (see \eqref{eq:phid0} and \eqref{eq:phid2})
\begin{align*}
c(z),s(w)>0,\qquad
c(w),s(z)<0.
\end{align*}
In this case, both the denominator and the numerator of \eqref{eq:addelliptic1a} have indeterminate sign.
More precisely, we have
\begin{align*}
&s(z)
=
-\left|s(z)\right|&&
-c(z)^2c(w)s(w)
=
\left|c(z)^2c(w)s(w)\right|\\
&c(z)c(w)^2
=
\left|c(z)c(w)^2\right|&&
-s(z)^2s(w)
=
-\left|s(z)^2s(w)\right|.
\end{align*}
The subtraction-free form is
\begin{align*}
{\left|c(z)c(w)^2\right|}
\left|c(z+w)\right|
+
\left|s(z)\right|
=
\left|s(z)^2s(w)\right|
\left|c(z+w)\right|
+
{\left|c(z)^2c(w)s(w)\right|}
\end{align*}
or
\begin{align*}
{\left|c(z)c(w)^2\right|}
\left|c(z+w)\right|
+
{\left|c(z)^2c(w)s(w)\right|}
=
\left|s(z)^2s(w)\right|
\left|c(z+w)\right|
+
\left|s(z)\right|.
\end{align*}
We then obtain the following in the limit $\varepsilon\to0$
\begin{align}
\max\left(
\tilde c(u)+2\tilde c(v)+\tilde c(u+v), 
\tilde s(u)
\right)
=
\max\left(
2\tilde s(u)+\tilde s(v)+\tilde c(u+v), 
2\tilde c(u)+\tilde c(v)+\tilde s(v)
\right).
\label{eq:udindet}
\end{align}
In general, the value of $\tilde c(u+v)$ can not be determined uniquely from those of $\tilde c(u)$, $\tilde c(v)$, $\tilde s(u)$, and $\tilde s(v)$ in terms of \eqref{eq:udindet}.
Thus we see that the case when $z\in d_1$ and $w\in d_3$ the addition formula for the ultradiscrete elliptic functions can not be reduced from \eqref{eq:addelliptic1a} through the ultradiscretization.

\begin{table}[htbp]
\renewcommand{\arraystretch}{1.2}
\begin{center}
\caption{The signs appearing in (\ref{eq:addelliptic1a} -- \ref{eq:addelliptic3b}).
The denominator and the numerator of the right hand side of each equation are denoted by ``d" and ``n" respectively. 
The symbol $\pm$ stands for indeterminate sign.}
\begin{tabular}{cccccccccccccccccc}\label{tab:udadd}
\\\HLINE
\multicolumn{2}{c}{Points}&
\multicolumn{4}{c}{Elliptic functions}&
\multicolumn{2}{c}{\eqref{eq:addelliptic1a}}&
\multicolumn{2}{c}{\eqref{eq:addelliptic1b}}&
\multicolumn{2}{c}{\eqref{eq:addelliptic2a}}&
\multicolumn{2}{c}{\eqref{eq:addelliptic2b}}&
\multicolumn{2}{c}{\eqref{eq:addelliptic3a}}&
\multicolumn{2}{c}{\eqref{eq:addelliptic3b}}\\
$z$&$w$&$c(z)$&$s(z)$&$c(w)$&$s(w)$&d&n&d&n&d&n&d&n&d&n&d&n\\\hline
$d_{1}$&$d_{1}$&
$+$&$-$&$+$&$-$&
$+$&${\pm}$&
$+$&$-$&
$-$&${\pm}$&
$-$&$+$&
$\color{red}{\pm}$&$\color{red}{\pm}$&
$\color{red}{\pm}$&$\color{red}{\pm}$\\
$d_{1}$&$d_{2}$&
$+$&$-$&$-$&$-$&
$+$&$-$&
$+$&${\pm}$&
$\color{red}{\pm}$&$\color{red}{\pm}$&
$\color{red}{\pm}$&$\color{red}{\pm}$&
$-$&$+$&
$-$&${\pm}$\\
$d_{1}$&$d_{3}$&
$+$&$-$&$-$&$+$&
$\color{red}{\pm}$&$\color{red}{\pm}$&
$\color{red}{\pm}$&$\color{red}{\pm}$&
${\pm}$&$-$&
${\pm}$&$+$&
${\pm}$&$+$&
${\pm}$&$-$\\
$d_{2}$&$d_{2}$&
$-$&$-$&$-$&$-$&
${\pm}$&$-$&
${\pm}$&$+$&
${\pm}$&$+$&
${\pm}$&$-$&
$\color{red}{\pm}$&$\color{red}{\pm}$&
$\color{red}{\pm}$&$\color{red}{\pm}$\\
$d_{2}$&$d_{3}$&
$-$&$-$&$-$&$+$&
$-$&${\pm}$&
$-$&$+$&
$\color{red}{\pm}$&$\color{red}{\pm}$&
$\color{red}{\pm}$&$\color{red}{\pm}$&
$+$&${\pm}$&
$+$&$-$\\
$d_{3}$&$d_{3}$&
$-$&$+$&$-$&$+$&
$-$&$+$&
$-$&${\pm}$&
$+$&$-$&
$+$&${\pm}$&
$\color{red}{\pm}$&$\color{red}{\pm}$&
$\color{red}{\pm}$&$\color{red}{\pm}$\\
\HLINE
\end{tabular}
\end{center}
\end{table}

This fact suggests that if both the denominator and the numerator have indeterminate signs then ordinary procedure of ultradiscretization
can not be applied\footnote{
Such a phenomenon is often referred as ``the problem of negativity" of the ultradiscretization \cite{IMNS06}.}; 
otherwise, we can apply it to the addition formulae (\ref{eq:addelliptic1a} -- \ref{eq:addelliptic3b}).
We summarize the signs of the equations (\ref{eq:addelliptic1a} -- \ref{eq:addelliptic3b}) for the choice of $z$ and $w$ in table \ref{tab:udadd}.
From table \ref{tab:udadd}, we observe that we can apply ordinary procedure of ultradiscretization to (\ref{eq:addelliptic1a} -- \ref{eq:addelliptic3b}) except for the following case
\begin{align*}
&z,w\in d_j
\quad
(j=1,2,3)
&\Longrightarrow&
&&\mbox{\eqref{eq:addelliptic3a} and \eqref{eq:addelliptic3b}}\\
&z\in d_j, w\in d_{j+1}
\quad
(j=1,2,3)
&\Longrightarrow&
&&\mbox{\eqref{eq:addelliptic2a} and \eqref{eq:addelliptic2b}}\\
&z\in d_j, w\in d_{j+2}
\quad
(j=1,2,3)
&\Longrightarrow&
&&\mbox{\eqref{eq:addelliptic1a} and \eqref{eq:addelliptic1b}},
\end{align*}
where the subscripts are reduced modulo 3.

Thus we have the following theorem.
\begin{theorem}\normalfont\label{thm:adformuellip}
Assume $u\in\overline{D_j}$ for a fixed $j=1,2,3$, where $\overline{D_j}$ is the closure of ${D_j}$.
Then the ultradiscrete elliptic functions $\tilde c$ and $\tilde s$ satisfy the following addition formulae
\begin{subequations}
\begin{align}
&\tilde c(u+v)
=
\max\left(
\tilde s(u),2\tilde c(u)+\tilde c(v)+\tilde s(v)
\right)
-
\max\left(
\tilde c(u)+2\tilde c(v),2\tilde s(u)+\tilde s(v)
\right)
\label{eq:uaddelliptic1a}\\
&\tilde s(u+v)
=
\max\left(
\tilde c(u)+\tilde s(u)+2\tilde s(v),\tilde c(v)
\right)
-
\max\left(
\tilde c(u)+2\tilde c(v),2\tilde s(u)+\tilde s(v)
\right),
\label{eq:uaddelliptic1b}
\end{align}
\end{subequations}
if and only if $v\in \overline{D_j\cup D_{j+1}}$, 
\begin{subequations}
\begin{align}
&\tilde c(u+v)
=
\max\left(
\tilde c(u)+\tilde s(u)+2\tilde c(v),\tilde s(v)
\right)
-
\max\left(
\tilde s(u)+2\tilde s(v),2\tilde c(u)+\tilde c(v)
\right)
\label{eq:uaddelliptic2a}\\
&\tilde s(u+v)
=
\max\left(
\tilde c(u),2\tilde s(u)+\tilde c(v)+\tilde s(v)
\right)
-
\max\left(
\tilde s(u)+2\tilde s(v),2\tilde c(u)+\tilde c(v)
\right),
\label{eq:uaddelliptic2b}
\end{align}
\end{subequations}
if and only if $v\in \overline{D_j\cup D_{j+2}}$, or
\begin{subequations}
\begin{align}
&\tilde c(u+v)
=
\max\left(
\tilde c(u)+2\tilde s(v),2\tilde s(u)+\tilde c(v)
\right)
-
\max\left(
\tilde c(u)+\tilde s(u),\tilde c(v)+\tilde s(v)
\right)
\label{eq:uaddelliptic3a}\\
&\tilde s(u+v)
=
\max\left(
\tilde s(u)+2\tilde c(v),2\tilde c(u)+\tilde s(v)
\right)
-
\max\left(
\tilde c(u)+\tilde s(u),\tilde c(v)+\tilde s(v)
\right),
\label{eq:uaddelliptic3b}
\end{align}
\end{subequations}
if and only if $v\in \overline{D_{j+1}\cup D_{j+2}}$, where the subscripts are reduced modulo 3.
\end{theorem}

(Proof)\quad
The `` if " part can be shown by using such limiting procedure as demonstrated above.
For the boundary values of the closures, $u_0$, $u_1$, and $u_2$, the formulae can be shown by direct calculation.
By substituting appropriate values, say $u\in D_1$ and $v\in D_{3}$, into \eqref{eq:uaddelliptic1a}, then we find that the equation does not hold.
In a similar manner, we can prove the `` only if " part for all cases. 
\qed

It immediately follows the addition formula for the points on the tropical Hesse curve $C_K$.
\begin{corollary}\normalfont
\label{cor:adformuellip}
Let $P=(X,Y)$ be a point on an edge $E_j$ of the tropical Hesse curve $C_K$ for a fixed $j=1,2,3$.
Then the point $P\uplus Q=(X\uplus X^\prime, Y\uplus Y^\prime)$ is given by the following addition formulae
\begin{subequations}
\begin{align}
&X\uplus X^\prime
=
\max\left(
Y,2X+X^\prime+Y^\prime
\right)
-
\max\left(
X+2X^\prime,2Y+Y^\prime
\right)
\label{eq:uaddHesse1a}\\
&Y\uplus Y^\prime
=
\max\left(
X+Y+2Y^\prime,X^\prime
\right)
-
\max\left(
X+2X^\prime,2Y+Y^\prime
\right),
\label{eq:uaddHesse1b}
\end{align}
\end{subequations}
if and only if $Q=(X^\prime,Y^\prime)\in E_j\cup E_{j+1}$, 
\begin{subequations}
\begin{align}
&X\uplus X^\prime
=
\max\left(
X+Y+2X^\prime,Y^\prime
\right)
-
\max\left(
Y+2Y^\prime,2X+X^\prime
\right)
\label{eq:uaddHesse2a}\\
&Y\uplus Y^\prime
=
\max\left(
X,2Y+X^\prime+Y^\prime
\right)
-
\max\left(
Y+2Y^\prime,2X+X^\prime
\right),
\label{eq:uaddHesse2b}
\end{align}
\end{subequations}
if and only if $Q\in E_j\cup E_{j+2}$, or
\begin{subequations}
\begin{align}
&X\uplus X^\prime
=
\max\left(
X+2Y^\prime,2Y+X^\prime
\right)
-
\max\left(
X+Y,X^\prime+Y^\prime
\right)
\label{eq:uaddHesse3a}\\
&Y\uplus Y^\prime
=
\max\left(
Y+2X^\prime,2X+Y^\prime
\right)
-
\max\left(
X+Y,X^\prime+Y^\prime
\right),
\label{eq:uaddHesse3b}
\end{align}
\end{subequations}
if and only if $Q\in E_{j+1}\cup E_{j+2}$, where the subscripts are reduced modulo 3.
\end{corollary}


\section{Conclusion}
We give the addition formula (\ref{eq:uaddHesse1a} -- \ref{eq:uaddHesse3b}) for the tropical Hesse pencil via the ultradiscretization of that (\ref{eq:addtheta1a} -- \ref{eq:addtheta3c}) for the level-three theta functions.
Each pair (\ref{eq:uaddHesse1a}, \ref{eq:uaddHesse1b}), (\ref{eq:uaddHesse2a}, \ref{eq:uaddHesse2b}), or (\ref{eq:uaddHesse3a}, \ref{eq:uaddHesse3b}) holds except for an edge of the curve, while those (\ref{eq:addtheta1a} -- \ref{eq:addtheta1c}), (\ref{eq:addtheta2a} -- \ref{eq:addtheta2c}), or (\ref{eq:addtheta3a} -- \ref{eq:addtheta3c}) holds except for three of the 9 zeros of the theta functions on $\C/L_\tau$.
In the tropical case, two of the three pairs are essentially the same where both of them are defined. 
Therefore, the addition formula uniquely determines the additive group structure of the tropical Hesse pencil in analogy to the original (non-tropical) case.

In \cite{KKNT09}, we construct the solvable chaotic dynamical system via the duplication formula for the tropical Hesse pencil.
The ultradiscrete QRT map $P=(X,Y)\mapsto\bar P=P\uplus T=(\bar X,\bar Y)$ can similarly be constructed by using the addition $\uplus$ of the tropical Hesse pencil.
For example, if we choose $V_3$ as $T$ then, by using corollary \ref{cor:adformuellip},  we obtain the linear map: 
\begin{align*}
&\bar X
=
-X+Y\\
&\bar Y
=
-Y+\bar X.
\end{align*}
This map is periodic with period three for any initial value because $V_3$ is the three-torsion point of the pencil.
This reflects the correspondence $\eta: p_6, p_7, p_8\mapsto V_3$ , where $p_6$, $p_7$, and $p_8$ are the three-torsion points of the Hesse pencil.
Thus we can construct both chaotic and integrable dynamical systems by using the group structure of the tropical Hesse pencil.

\section*{Acknowledgment}
The author would like to express his sincere thanks to Professor Kenji Kajiwara for fruitful discussion.
This work was partially supported by grants-in-aid for scientific research, Japan society for the promotion of science (JSPS) 19740086 and 22740100.

\end{document}